\documentclass[12pt]{article}
\usepackage[english,activeacute]{babel}
\usepackage{natbib}
\usepackage{comment}
\usepackage{float}
\usepackage[hidelinks]{hyperref}
\usepackage{mathrsfs}
\usepackage{enumitem}
\usepackage[font={small,it}]{caption}
\usepackage{amsmath,amsfonts,amsthm,amssymb}
\usepackage{bm,rotating,multirow,dsfont,graphicx}
\usepackage[usenames, dvipsnames]{color}
\usepackage{url}
\usepackage{multicol}
\usepackage{multirow}
\usepackage[T1]{fontenc}
\usepackage{flafter}
\usepackage{appendix}
\usepackage{subfigure}
\usepackage{xcolor}
\usepackage{soul}
\makeatletter
\def\hlinewd#1{%
	\noalign{\ifnum0=`}\fi\hrule \@height #1 %
	\futurelet\reserved@a\@xhline}
\makeatother
\def\spacingset#1{\renewcommand{\baselinestretch}{#1}\small\normalsize}\spacingset{1}
\def\@roman#1{\romannumeral #1}
\begin{document}

\title{Stochastic Gradient Variational Bayes in the \\ Stochastic Blockmodel}

\date{}

\author{
    Pedro Regueiro, Nubank, Mexico \\
    Abel Rodríguez, University of Washington, US \\
    Juan Sosa, Universidad Nacional de Colombia, Colombia\footnote{Corresponding author: jcsosam@unal.edu.co.}
}

\maketitle

\begin{abstract} 
Stochastic variational Bayes algorithms \citep{Jordanetal99,Jaakkola&Jordan00} have become very popular in the machine learning literature, particularly in the context of nonparametric Bayesian inference.  These algorithms replace the true but intractable posterior distribution with the best (in the sense of Kullback-Leibler divergence) member of a tractable family of distributions, using stochastic gradient algorithms to perform the optimization step.  stochastic variational Bayes inference implicitly trades off computational speed for accuracy, but the loss of accuracy is highly model (and even dataset) specific.  In this paper we carry out an empirical evaluation of this trade off in the context of stochastic blockmodels, which are a widely used class of probabilistic models for network and relational data.  Our experiments indicate that, in the context of stochastic blockmodels, relatively large subsamples are required for these algorithms to find accurate approximations of the posterior, and that even then  the quality of the approximations provided by stochastic gradient variational algorithms can be highly variable.  
\end{abstract}


\spacingset{1.1} 

\newpage

\section{Introduction}\label{intro}

The main bottleneck in the application of Bayesian statistical methods to large-scale data is computational:  Standard tools for performing high-dimensional numerical integration such as Markov chain Monte Carlo (MCMC) algorithms \citep{robertcasella2004monte} typically scale poorly with the number of observations in the sample.  In the context of maximum likelihood inference, stochastic gradient algorithms (e.g., see \citealp{DBLP:journals/corr/Ruder16} for a review) have become extremely popular in recent years.  These approaches, which trace their roots to the work of \cite{Robbins&Monro51} and \cite{kiefer1952stochastic}, are first-order optimization methods that replace the true gradient of the likelihood function with a noisy version computed over a small subset of observations (often a single one).  More recently, work on second-order stochastic gradient methods has started to become popular (e.g., see \citealp{agarwal2017second}).

Stochastic gradient methods have been used in the context of Bayesian inference to develop Hamiltonian Monte Carlo algorithms that rely on stochastic gradients (e.g., see \citealp{welling2011bayesian,chen2014stochastic,ma2015complete}), or variational approximations in which the underlying optimization is performed using stochastic gradient methods \citep{Hoffmanetal13}.  Of these, stochastic gradient variational Bayes (SGVB) approximations, which replace the intractable posterior distribution with a tractable approximation whose parameters are chosen to minimize the Kullback-Leibler divergence between them, have become particularly popular (e.g., see \citealp{Fouldsetal13,Kimetal13,kingma2014stochastic,knowles2015stochastic,Polatkanetal15,hoffman2015structured,mandt2016variational,tran2021variational,kim2022markov,cao2023variational,mcnamara2024sequential}).

Variational Bayes inference can be extremely powerful, but the properties of the algorithm are not well understood \citep{blei2017variational} and its performance depends heavily on the class of models being approximated, the specific dataset being analyzed, and the parameters of interest.  Theoretical results supporting the use of variational approximations for Bayesian inference are limited.  Some notable exceptions include \cite{wang2004lack,wang2004convergence}, \cite{Celisseetal12} and \cite{Bickeletal13}, all of which are concerned with the properties of point estimators rather than the full posterior distribution.  Results associated with stochastic gradient versions of variational inference are even more scarce.  Empirical evidence supporting (or discouraging) the use of variational methods in specific settings is particularly limited, as few papers present detailed comparisons against alternatives such as MCMC algorithms.

The computational challenges associated with the Bayesian analysis of large datasets are particularly acute in the context of network data, where the computational complexity of most algorithms tends to increase as the square of the number of nodes in the network.  This high computational complexity means that problems with even a moderate number of nodes might be intractable.  This is an interesting challenge since network data has increasingly become the focus of interest in statistics and computer science.  Examples of statistical models for network data that are widely used in practice include  the configuration model \citep{Bender&Canfield78}, exponential random graph models \citep{Frank&Strauss86}, latent social space models \citep{Hoffetal02,hoff2005bilinear,hoff-2008,hoff-2009-multiplicative}, and the stochastic blockmodel \citep{Hollandetal83}.  See \cite{Goldenbergetal10} for a review.

In this paper, we are particularly interested in statistical inferences under stochastic blockmodels, which extends the ideas behind clustering methods to network data by dividing the nodes of the network into ``structurally equivalent'' groups.   Algorithms that aim at generation partitions of this kind are sometimes referred to as ``community detection'' algorithms in the literature (e.g., see \citealp{Newman04,Schaeffer07,Porteretal09,Fortunato10} and \citealp{li2024comprehensive} for a comprehensive review).  Key advantages of stochastic blockmodels in the context of community detection are their ability to provide measures of uncertainty on the partition structure, handle missing observations, and identify disassortative communities.

Variational algorithms have been used in the context of stochastic blockmodels and its extensions (e.g., see \citealp{Airoldietal08,mariadassou2010uncovering,latouche2012variational,Gopalanetal12,tabouy2020variational}).  However, as far as we are aware, there has been no thorough empirical evaluation of the stochastic gradient version of the algorithm, specially \textit{vis-a-vis} established techniques such as MCMC.  For example, it is well-known that posterior distributions for stochastic blockmodels can be highly multimodal.  This can represent a problem even for MCMC algorithms, but the situation is potentially worse in the case of variational inference because the underlying optimization problem is itself multimodal.  Furthermore, SGVB presents specific challenges in the context of stochastic blockmodels, such as the selection of the minibatch structure and size.

The aim of this paper is to empirically evaluate the performance of SGVB for stochastic blockmodels. First, by benchmarking it against MCMC algorithms in situations with a moderate number of nodes where it is possible using both algorithms, and then, by testing it in real as well as simulated high-dimensional datasets.  Although the stochastic gradient variational (SGV) approximation have been developed for a number of extensions of the stochastic blockmodel, we work with the simplest version of model in order to provide an upper bound of the performance of the algorithm.  The results from these comparisons can provide guidance into the practical limitations of SGVB algorithms in the context of stochastic blockmodels, and the reliability of the conclusions derived from analyses that rely on them.  Some of our findings include:  (1) the performance of SGVB is relatively robust to the choice of the parameters controlling the learning rate of the stochastic gradient algorithm, (2) although SGVB algorithms are extremely fast, the size of the minibatch required to get reasonable approximations to the posterior distribution is often relatively large, which limits the computational gains derived from the algorithm, (3) in networks with a large number of small communities, SGVB can produce very poor approximations to the posterior distribution, resulting in a very poor predictive performance.

The remaining of this paper is organized as follows:  Section 2 reviews classical Bayesian approaches to inference in stochastic blockmodels, including MCMC algorithms.  Section 3 reviews variational approximations and their stochastic gradient variants, both in general and in the context of stochastic blockmodels.  Section 4 evaluates the performance of SGVB algorithms for stochastic blockmodels in several simulated and real datasets.  This evaluation includes comparisons with MCMC algorithms in small and medium sized datasets where such comparisons are viable.  Finally, Section 5 provides a discussion of our findings.

\section{Stochastic blockmodels}\label{bb}

In this paper, we focus on algorithms for stochastic blockmodels for undirected binary network data with adjacency matrix 
\begin{equation*}
y_{i,j} = 
\begin{cases} 
1 \quad &\text{if there is an edge from vertex $i$ to vertex $j$;} \\ 
0 \quad &\text{otherwise,} 
\end{cases}
\end{equation*}
where $I$ is the number of vertices in the network \citep{Hollandetal83,wang1987stochastic,snijders1997estimation}. For any undirected network the adjacency matrix is, by construction, symmetric, and therefore, it is possible to disregard the observations below (or above) the main diagonal.  Furthermore, in the case of a network without self interactions, it is also possible to disregard the observations on the main diagonal as they are all assumed to be structural zeros.  Consequently, the set of observations is taken to be
$$
\mathcal{Y}=\{y_{i,j}: 1\leq i<j\leq I,\; i,j\in\mathbb{N}\}.
$$

The binary nature of the interactions naturally suggests a Bernoulli distribution,
\begin{equation}\label{lm1}
y_{i,j}\mid \lambda_{i,j} \sim \textsf{Ber}(\lambda_{i,j}), \qquad 1\leq i<j\leq I.
\end{equation}
The basic idea behind stochastic blockmodels is that individuals can be clustered into $K \le I$ unknown blocks.  These blocks corresponds to groups of structurally equivalent nodes, i.e., nodes that have similar relationships with other nodes. Hence, the probability of observing an interaction between two individuals is modelled as a function of their respective blocks,
\begin{equation*}
\lambda_{i,j} = \Pr \left( y_{i,j} = 1 \mid \xi_i, \xi_j, \{\theta_{k,l}\}_{k,l=1}^K \right) = \theta_{\xi_i,\xi_j}\,,
\end{equation*}
where the latent variables $\xi_1,\ldots,\xi_I$ are block indicators, and the elements of $\{\theta_{k,l}\}_{k,l=1}^K$ are usually referred to as community parameters.  Notice that, again, because of the inherent symmetry restriction, attention can be constraint to
\begin{equation*}
\Theta=\{\theta_{k,l}: 1\leq k\leq l \leq K,\; k,l\in\mathbb{N}\},
\end{equation*}
so that, the model can be written as 
$$
\Pr \left( y_{i,j} = 1 \mid \xi_i, \xi_j, \Theta \right) = \theta_{\min\{\xi_i,\xi_j \}, \max\{\xi_i,\xi_j \}}.
$$

Assuming conditional independence across both interactions and actors, the likelihood function can be expressed as the product 
\begin{equation}\label{likeasprod}
p(\mathcal{Y}\mid \Theta,\boldsymbol{\xi})=\prod_{i=1}^{I-1}\prod_{j=i+1}^I p\left(y_{i,j}\mid \theta_{\phi\left(\xi_i,\xi_j\right)}\right).
\end{equation}
where $\boldsymbol{\xi} = (\xi_1,\ldots,\xi_I)$ and $\phi(u,v)=(\min\{u,v\},\max\{u,v\})$.  In turn, \eqref{likeasprod} can be rewritten as 
\begin{align}
p(\mathcal{Y}\mid \Theta,\boldsymbol{\xi}) &= \prod_{k=1}^K\prod_{l=k}^K \prod_{i=1}^{I-1}\prod_{j=i+1}^I  \left\{ \theta_{k,l}^{y_{i,j}} \left(1-\theta_{k,l}\right)^{1-y_{i,j}}  \right\}^{\mathsf{1} ( (k,l) = \phi(\xi_i, \xi_j) )}  \nonumber \\
& = \prod_{k=1}^K\prod_{l=k}^K\theta_{k,l}^{s_{k,l}} \left(1-\theta_{k,l}\right)^{n_{k,l}-s_{k,l}} , \label{likeprodgen}
\end{align}
with $1(\cdot)$ representing the indicator function, $s_{k,l}=\sum_{\mathcal{S}_{k,l}}y_{i,j}$, $n_{k,l}=\sum_{\mathcal{S}_{k,l}}1$, and the sum is taken over the set 
\begin{equation*}
\mathcal{S}_{k,l}=\left\{(i,j):\; i<j,\; (k,l)=\phi\left(\xi_i,\xi_j\right)\right\}.
\end{equation*}

Bayesian inference for the stochastic blockmodel requires the specification of priors for the unknown parameters $\boldsymbol{\xi}$ and $\Theta$.  On the one hand, community interaction parameters are often conveniently modeled as independent and identically distributed a priori from a Beta distribution,  
\begin{equation}\label{betaprior}
\theta_{k,l}\sim\textsf{Beta}(a,b).
\end{equation}
On the other hand, for the block indicators it is common to use a hierarchical prior specification where $\Pr(\xi_i = k) = w_k$ independently for each $k$ and $\boldsymbol{w}=(w_1, \ldots, w_K)$ is assigned a Dirichlet prior.  This is the approach of \cite{snijders1997estimation}.

An alternative prior specification that allows for a flexible estimation of the number of blocks in the network is the \emph{infinite relational model} of \cite{Kempetal06}. This model allows for the effective number of communities in the network $K^\star \leq K = \infty$ to be learned from the data. Specifically, $\boldsymbol{\xi}$ is assumed to follow a Chinese restaurant process (CRP) prior, which implies that its distribution is given by the Ewens sampling formula \citep{Ewens72},
\begin{equation*}
p(\boldsymbol{\xi})=\frac{\Gamma(\alpha)\,\alpha^{K^\star}}{\Gamma(\alpha+I)}\prod_{k=1}^{K^\star}\Gamma(m_k),
\end{equation*}
where $m_k$ is the size of group $k$ and $\alpha$ is a concentration parameter. This parameter in the infinite relational model controls the number of realized communities $K^\star$.  In particular, \cite{Antoniak74} shows that the distribution of $K^{\star}$ is given by
\begin{equation*}
\Pr(K^\star=k\mid\alpha)=S(I,k)\alpha^k\frac{\Gamma(\alpha)}{\Gamma(\alpha+I)} ,
\end{equation*}
where $S(\cdot, \cdot)$ represents the unsigned Stirling numbers of the first kind \citep{graham1994concrete}.

In this paper, we focus on the finite-dimensional Dirichlet-multinomial prior with
$$
\boldsymbol{w} \sim \textsf{Dir} \left(\frac{\alpha}{K},\frac{\alpha}{K},\ldots,\frac{\alpha}{K}\right),
$$
which approximates the infinite relational model with concentration parameter $\alpha$ for large $K$ (e.g., see \citealp{Ishwaran&Zarepour00} and \citealp{Neal00}).  The hyperparameters $\alpha,$ $a$ and $b$ can be fixed, or alternatively, an additional hierarchical layer can be added by incorporating prior distributions $\pi(\alpha)$ and $\pi(a,b)$.  In the sequel we treat the hyperparameters as constants and set $a=b=1$, i.e., we use a uniform prior on the interaction probabilities.

\subsection{Markov chain Monte Carlo computation}

The model described above leads to a posterior distribution of the form
\begin{equation} \label{postbeta}
p(\Theta,\boldsymbol{\xi},\boldsymbol{w}\mid \mathcal{Y}) \propto \prod_{k=1}^K\prod_{l=k}^K\left(\theta_{k,l}\right)^{a+s_{k,l}-1}\left(1-\theta_{k,l}\right)^{b+n_{k,l}-s_{k,l}-1}\prod_{k=1}^K w_k^{\frac{\alpha}{K}+m_k-1},
\end{equation}
where $m_k=\sum_{\mathcal{S}_k}1$ and the sum is taken over $\mathcal{S}_k=\{i: \xi_i=k\}$.  This posterior distribution is not analytically tractable.  Markov Chain Monte Carlo (MCMC) algorithms \citep{Metropolisetal53, Hastings70, Geman&Geman84, Gelfand&Smith90, gamerman2006markov} are a widely used class of computational algorithms used to approximate summaries of such posterior distribution.  MCMC algorithms rely on sampling methods to approximate the posterior distribution. Here, we briefly describe an MCMC algorithm for the stochastic blockmodel described above.

Note that the full conditional distribution for each indicator variable $\xi_i$ is given by
\begin{equation}\label{fcxibeta}
\Pr(\xi_{i}=k\mid \Theta,\boldsymbol{\xi}_{-i},\boldsymbol{w},\mathcal{Y})\propto w_{k}\prod_{\substack{j=1 \\ j\neq i}}^I\left(\theta_{\phi\left(\xi_j,k\right)}\right)^{y_{\phi\left(i,j\right)}}\left(1-\theta_{\phi\left(\xi_j,k\right)}\right)^{1-y_{\phi\left(i,j\right)}},
 \end{equation}
for $k\in\{1,\ldots,K\}$.  Here, $\boldsymbol{\xi}_{-i}$ represents the vector $\boldsymbol{\xi}$ with its $i$-th component removed.  Thus, for every $i=1,\ldots,I$, $\xi_{i}$ can be sampled from a Categorical distribution with weights vector given by normalizing the RHS of equation \eqref{fcxibeta}.  Now, in the case of $\Theta$, the entries are conditionally independent of each other with
\begin{equation}
p(\theta_{k,l}\mid ,\boldsymbol{\xi},\boldsymbol{w},\mathcal{Y}) \propto \left(\theta_{k,l}\right)^{a+s_{k,l}-1}\left(1-\theta_{k,l}\right)^{b+n_{k,l}-s_{k,l}-1} .
\end{equation}
This is identified as the kernel of a Beta distribution with parameters $a+s_{k,l}$ and $b+n_{k,l}-s_{k,l}$.  Note that, in the case of an empty component (i.e., $m_k = 0$), the full conditional distribution reduces to the prior.  Finally, for the weights,
\begin{equation}\label{wfcbp}
p(\boldsymbol{w}\mid \Theta,\boldsymbol{\xi},\mathcal{Y}) \propto \prod_{k=1}^K w_k^{\frac{\alpha}{K}+m_k-1} ,
\end{equation}
which corresponds to the kernel of a Dirichlet distribution with parameter vector $$
\left(\frac{\alpha}{K}+m_1,\ldots,\frac{\alpha}{K}+m_K\right).
$$

\section{Variational approximations}\label{va}

The MCMC algorithm we just described is relatively easy to implement, and the Markov chain implicitly defined is guaranteed to eventually converge to the posterior distribution, which in principle allows us to control the accuracy of the approximation by adjusting the number of samples generated.  In practice, however, as the number of nodes grows, the computational burden can make this approach infeasible.  Specific to the setting of network analysis, notice that as the number of nodes $I$ grows, the number of interactions grows as $\mathcal{O}(I^2)$.  Similarly, the number of communities might be expected to grow as the number of observations increases, increasing the number of parameters that need to be estimated.  Thus, MCMC algorithms can be impractical, even for moderately large networks.

Variational approximations (e.g., see \citealp{Jordanetal99} and \citealp{Jaakkola&Jordan00}) are an alternative to MCMC algorithms that easily scale up to models with large numbers of observations and parameters.  In general, consider the problem of approximating an unknown function ($p$) that can be evaluated up to a proportionality constant, by another function ($q$) that is restricted to be a member of a certain family of functions.  To this end, it is possible to define a functional measure of ``dissimilarity'' between $p$ and $q$, and use calculus of variations techniques to minimize that measure, thus finding $q$ in such family that is ``closest'' to $p$.  The idea of applying this technique to the case where $p$ is chosen to be a posterior distribution can be traced back to the mid 90's, in works such as \cite{Sauletal96}, and it is nowadays known as \textit{variational Bayes} in the specialized literature.

Briefly, the main idea can be summarized as follows. Let $\boldsymbol{\varphi}$ be a set of parameters and $p(\boldsymbol{\varphi}\mid\mathbf{x})$ its posterior distribution after data $\mathbf{x}$ has been observed.  The purpose is to approximate  $p(\boldsymbol{\varphi}\mid\mathbf{x})$ with $q(\boldsymbol{\varphi})$. Specifically, if the Kullback--Leibler divergence is chosen as a measure of dissimilarity, the problem becomes 
 \begin{equation}\label{vbproblem}
 \underset{q}{\min} \int q(\boldsymbol{\varphi}) \log \frac{q(\boldsymbol{\varphi})}{p(\boldsymbol{\varphi}\mid\mathbf{x})}\, \textsf{d}\boldsymbol{\varphi}.
\end{equation}

Now, it can be easily shown that  
\begin{equation*}
 \int q(\boldsymbol{\varphi}) \log \frac{q(\boldsymbol{\varphi})}{p(\boldsymbol{\varphi}\mid\mathbf{x})} \,\textsf{d}\boldsymbol{\varphi}=\log p(\mathbf{x}) -  \int q(\boldsymbol{\varphi}) \log \frac{p(\boldsymbol{\varphi},\mathbf{x})}{q(\boldsymbol{\varphi})} \,\textsf{d}\boldsymbol{\varphi}
\end{equation*} 
and, therefore, minimizing the Kullback--Leibler divergence of $q$ with respect to $p$ is equivalent to maximizing $\mathbb{E}_{q(\boldsymbol{\varphi})}\left[\log \frac{p(\boldsymbol{\varphi},\mathbf{x})}{q(\boldsymbol{\varphi})}\right]$, which is known in Physics as \emph{free energy}, and in Computer Science as \emph{evidence lower bound} (ELBO).  Note that the ELBO can be decomposed as 
\begin{equation*}
F(q,\mathbf{x})=\mathbb{E}_{q(\boldsymbol{\varphi})}[\log p(\mathbf{x},\boldsymbol{\varphi})]+H\left[q(\boldsymbol{\varphi})\right]
\end{equation*}
where $H$ denotes the Shannon entropy \citep{lesne2014shannon}.

It is common to assume that $q$ is restricted to satisfy the \textit{mean field assumption},
  \begin{equation*}
  q(\boldsymbol{\varphi})=\prod_{i} q_i(\varphi_{i}) ,
\end{equation*}
where the $q_i(\varphi_{i})$ are the marginal variational densities. The solution of this problem satisfies
 \begin{equation}\label{vbgensolution}
 \log q^\star_i(\varphi_{i})\propto\mathbb{E}_{q(\boldsymbol{\varphi_{-i}})}\left[ \log p(\boldsymbol{\varphi},\mathbf{x}) \right]  ,
\end{equation}
which leads to a coordinate optimization algorithm. In particular, when the full conditional posterior distributions associated with $p$ belong to the exponential family of distributions, the solution to \eqref{vbgensolution} is readily available.  Convergence of the algorithm is usually assessed through the value of the ELBO, with the algorithm being stopped when the relative change in the ELBO across iterations falls bellow a given threshold.  The ELBO is also used to compare solutions generated by different initial conditions.

\subsection{Variational inference for the stochastic blockmodel}

Consider the blockmodel with fixed hyperparameters as discussed in Section \ref{bb}. Before going into the variational algorithm, notice that when the interest lies in the community indicators only, it is possible to marginalize over the weight parameters to obtain 
\begin{equation}
p(\boldsymbol{\xi})= \frac{\Gamma(\alpha)}{\left[\Gamma\left(\frac{\alpha}{K}\right)\right]^K \Gamma(I+\alpha)}\prod_{k=1}^K\Gamma\left(\frac{\alpha}{K}+m_k\right)
\end{equation}
and, thus, 
\begin{equation}
\Pr(\xi_i=k\mid\boldsymbol{\xi}_{-i})=\frac{m_k^{-i}+\frac{\alpha}{K}}{(I-1)+\alpha} \; \text{for all } i\in\{1,\ldots,I\} \, \text{and all } k\in\{1,\ldots,K\},  
\end{equation} 
where $m_k^{-i}=\sum_{j\neq i} 1(\xi_j=k)$.

In this way, it is possible to find the variational approximation $q(\Theta,\boldsymbol\xi)$ to the marginal posterior distribution $p(\Theta,\boldsymbol\xi\mid\mathcal{Y})\propto p(\mathcal{Y}\mid\Theta,\boldsymbol{\xi})\,p(\Theta)\,p(\boldsymbol{\xi})$ rather than approximating the complete posterior distribution $p(\Theta,\boldsymbol\xi,\boldsymbol{w}\mid\mathcal{Y}).$  This, effectively reduces the number of constrains imposed by the mean field assumptions, and therefore, potentially improves the approximation. Now, for $1\leq k\leq l\leq K,$ the solution to \eqref{vbproblem} satisfies
\begin{equation*}
\log q^\star(\theta_{k,l})\propto\mathbb{E}_{q(\Theta_{-kl},\boldsymbol\xi)} [\log p(\mathcal{Y},\Theta,\boldsymbol{\xi})]\propto\mathbb{E}_{q(\Theta_{-kl},\boldsymbol\xi)} [\log p(\mathcal{Y}\mid\Theta,\boldsymbol{\xi})]+ \log p(\theta_{k,l}), 
\end{equation*}
which implies that 
\begin{equation}
q^\star(\theta_{k,l})\propto \exp\left\{\sum_{i=1}^{I-1}\sum_{j=i+1}^I [y_{i,j}\log\theta_{k,l}+(1-y_{i,j})\log(1-\theta_{k,l})]r_{k,l}^{i,j}\right\}\theta_{k,l}^{a-1}(1-\theta_{k,l})^{b-1},
\end{equation}
where 
\begin{equation*}
r_{k,l}^{i,j}=
\begin{cases}
q(\xi_i=k)q(\xi_j=l)+q(\xi_i=l)q(\xi_j=k)\quad &\text{ if } k\neq l;\\ q(\xi_i=k)q(\xi_j=l) \quad&\text{ if }k=l. 
\end{cases}
\end{equation*} 
That is, the variational distribution of $\theta_{k,l}$ is $\textsf{Beta}(a_{k,l}^\star,b_{k,l}^\star)$, with
\begin{equation}\label{globalvbupdates}
a_{k,l}^\star=a+\sum_{i=1}^{I-1}\sum_{j=i+1}^Ir_{k,l}^{i,j}y_{i,j}\quad\text{and}\quad b_{k,l}^\star=b+\sum_{i=1}^{I-1}\sum_{j=i+1}^Ir_{k,l}^{i,j}(1-y_{i,j}).
\end{equation}
Also, for $i \in \{1,2,\ldots,I\},$ it follows that
\begin{align*}
\log q^\star(\xi_{i}) & \propto\mathbb{E}_{q(\Theta,\boldsymbol{\xi_{-i}})} [\log p(\mathcal{Y},\Theta,\boldsymbol{\xi})] \\
                              & \propto\mathbb{E}_{q(\Theta,\boldsymbol{\xi_{-i}})} [\log p(\mathcal{Y}\mid\Theta,\boldsymbol{\xi})]+ \mathbb{E}_{q(\boldsymbol{\xi_{-i}})} [\log p(\xi_i\mid\boldsymbol{\xi}_{-i})].
\end{align*}

Following \cite{Kuriharaetal07}, the second term in the last expression can be approximated using the second order Delta method and the fact that $n^{-i}_{\xi_i}$ is a sum of independent Bernoulli random variables. Specifically, for any $i$
\begin{equation*}
\mathbb{E}_{q(\boldsymbol{\xi_{-i}})}\left[\log \left(n^{-i}_{\xi_i}+\frac{\alpha}{K}\right)\right]\approx\log\left( \sum_{j\neq i}q(\xi_j=\xi_i)+\frac{\alpha}{K}\right)-\frac{\displaystyle\sum_{j\neq i}q(\xi_j=\xi_i)(1-q(\xi_j=\xi_i))}{2\left[\displaystyle\sum_{j\neq i}q(\xi_j=\xi_i)\right]^2}.
\end{equation*}
Therefore, for any $k \in \{1,2,\ldots,K\}$,
\begin{multline}\label{localvbupdate}
\log q^\star(\xi_{i}=k) \propto  \sum_{j\neq i}\left[y_{\phi\left(i,j\right)} \chi_{k,1}+(1-y_{\phi\left(i,j\right)}) \chi_{k,2} \right] \\
+\log\left( \sum_{j\neq i}q(\xi_j=k)+\frac{\alpha}{K}\right)-\frac{1}{2}\frac{\sum_{j\neq i}q(\xi_j=k)(1-q(\xi_j=k))}{\left[\sum_{j\neq i}q(\xi_j=k)\right]^2},
\end{multline}
where 
\begin{gather*}
\chi_{k,1}=\sum_{l=1}^K\left[\left(\psi\left(a^\star_{\phi\left(k,l\right)}\right)-\psi\left(a^\star_{\phi\left(k,l\right)}+b^\star_{\phi\left(k,l\right)}\right)\right)q(\xi_j=l)\right],\\
\chi_{k,2}=\sum_{l=1}^K\left[\left(\psi\left(b^\star_{\phi\left(k,l\right)}\right)-\psi\left(a^\star_{\phi\left(k,l\right)}+b^\star_{\phi\left(k,l\right)}\right)\right)q(\xi_j=l)\right],
\end{gather*}
and $\psi(\cdot)$ represents the Digamma function. Finally, in a similar fashion, the free energy is given by 
\begin{align*}
F(q,\mathcal{Y}) &\approx
\frac{1}{2}K(K+1)\left(\log\Gamma(a+b)-\log\Gamma(a)-\log\Gamma(b)\right)+\log\Gamma(\alpha)-\log\Gamma(I+\alpha)\\
&\hspace{1cm}
-K\log\Gamma\left(\frac{\alpha}{K}\right)+\sum_{k=1}^K\sum_{l=k}^K \log\Gamma(a^\star_{k,l})+ \log\Gamma(b^\star_{k,l})- \log\Gamma(a^\star_{k,l}+b^\star_{k,l}) \\
&\hspace{2cm}
\sum_{k=1}^K\left[\log\Gamma\left(\sum_{i=1}^Iq(\xi_i=k)+\frac{\alpha}{K}\right)+ \right.\\
&\hspace{3cm}
\left.\psi_1\left(\sum_{i=1}^Iq(\xi_i=k)+\frac{\alpha}{K}\right)\left(\sum_{i=1}^Iq(\xi_i=k)(1-q(\xi_i=k))\right)\right]\\
&\hspace{4cm}
\sum_{i=1}^I\sum_{k=1}^K q(\xi_i=k)\log q(\xi_i=k),
\end{align*}
where $\psi_1(\cdot)$ denotes the Trigamma function.

\subsection{Stochastic variational inference for the stochastic blockmodel}\label{se:sgvb}

In general, variational Bayes algorithms tend to converge to a local optimum in a relatively small number of iterations, making them fast in comparison to MCMC algorithms.  However, because the calculations required to obtain $\mathbb{E}_{q(\boldsymbol{\varphi_{-i}})}\left[ \log p(\boldsymbol{\varphi},\mathbf{X}) \right]$ are similar to those of the corresponding full conditional $p(\varphi_i\mid\boldsymbol{\varphi_{-i}},\mathbf{X})$, these methods also suffer from scalability issues.  Essentially, the difficulty lies on the fact that, for each parameter, calculations involving the entire data matrix are required.

An extension of this algorithm introduced by \cite{Hoffmanetal13} to address these scalability concerns is the \emph{stochastic variational inference}.  Using stochastic optimization ideas, this algorithm speeds computations producing noisy estimates of the (natural) gradient within a coordinate ascent approach. This algorithm relies on a split of the parameter set into local and global parameters $\boldsymbol{\varphi}=(\boldsymbol{\varphi_l},\boldsymbol{\varphi_g})$ in such a way that $p(\boldsymbol{\varphi_l},\mathbf{x}\mid\boldsymbol{\varphi_g})=\prod_{i}p(\varphi_i,x_i\mid\boldsymbol{\varphi_g}),$  and $p(\boldsymbol{\varphi_l},\mathbf{x}\mid\boldsymbol{\varphi_g})$ and $p(\boldsymbol{\varphi_g})$ are conjugate in the exponential family.   Specifically, if 
   \begin{equation*}
p(\varphi_i,x_i\mid\boldsymbol{\varphi_g})=h(\varphi_i,x_i)\exp\{\boldsymbol{\varphi_g}^Tt(\varphi_i,x_i)-a_l(\boldsymbol{\varphi_g})\}
\end{equation*}
 and 
  \begin{equation*}
p(\boldsymbol{\varphi_g})= h(\boldsymbol{\varphi_g})\exp\{\mathbf{h}^Tt(\boldsymbol{\varphi_g})-a_g(\mathbf{h})\},
\end{equation*}
conjugacy implies that $t(\boldsymbol{\varphi_g})^T=(\boldsymbol{\varphi_g},-a_l(\boldsymbol{\varphi_g}))$, and thus,
  \begin{equation*}
p(\boldsymbol{\varphi_g}\mid\boldsymbol{\varphi_l},\mathbf{X})\propto h(\boldsymbol{\varphi_g})\exp\{\boldsymbol{\eta}^T(\boldsymbol{\varphi_l},\mathbf{X})t(\boldsymbol{\varphi_g})\},
\end{equation*}
where $\boldsymbol{\eta}^T(\boldsymbol{\varphi_l},\mathbf{x})=\mathbf{h}^T+(\sum_i^N t(\varphi_i,x_i), N),$ with $N$ the dimension of $\mathbf{x}.$
Furthermore, in that case the solution to the variational distribution is in the same exponential family 
\begin{equation*}
q(\boldsymbol{\varphi_g})\propto h(\boldsymbol{\varphi_g})\exp\{\mathbf{h_q}^Tt(\boldsymbol{\varphi_g})-a_g(\mathbf{h_q})\},
\end{equation*}
and therefore, the natural gradient of the free energy can be written as  
\begin{equation*}
\hat{\nabla}_\mathbf{h_q}F(q,\mathbf{x})=\mathbb{E}_{q^\star(\boldsymbol{\varphi_l})}[\boldsymbol{\eta}^T(\boldsymbol{\varphi_l},\mathbf{x})]-\mathbf{h_q},
\end{equation*}
where $q^\star(\boldsymbol{\varphi_l})$ is the regular variational distribution for the local parameters. This result can be used in a \cite{Robbins&Monro51} algorithm taking $\mathbf{x}^{(L)}$ a random subsample of the data, $\boldsymbol{\varphi_l}^{(L)}$ the corresponding local parameters and $\mathcal{C}$ the appropriate scaling factor, and setting  
 \begin{equation}\label{svbvupdate}
\mathbf{h_q}^{(t)}=(1-\rho_t)\mathbf{h_q}^{(t-1)}+\rho_t\mathcal{C}\mathbb{E}_{q^\star(\boldsymbol{\varphi_l^{(L)}})}[\boldsymbol{\eta}^T(\boldsymbol{\varphi_l}^{(L)},\mathbf{x}^{(L)})],
\end{equation}
which is guaranteed to converge to a local minimum as long as the positive real sequence $\rho_t$ satisfies
\begin{equation}\label{conditionsrho}
\sum_{t=1}^\infty \rho_t\rightarrow\infty \quad \text{and}\quad \sum_{t=1}^\infty \rho_t^2<\infty.
\end{equation}

\begin{figure}[!htb]
\vspace{-.5cm}
\centering
\begin{tabular}{@{}cc@{}}
    \includegraphics[width=.4\textwidth]{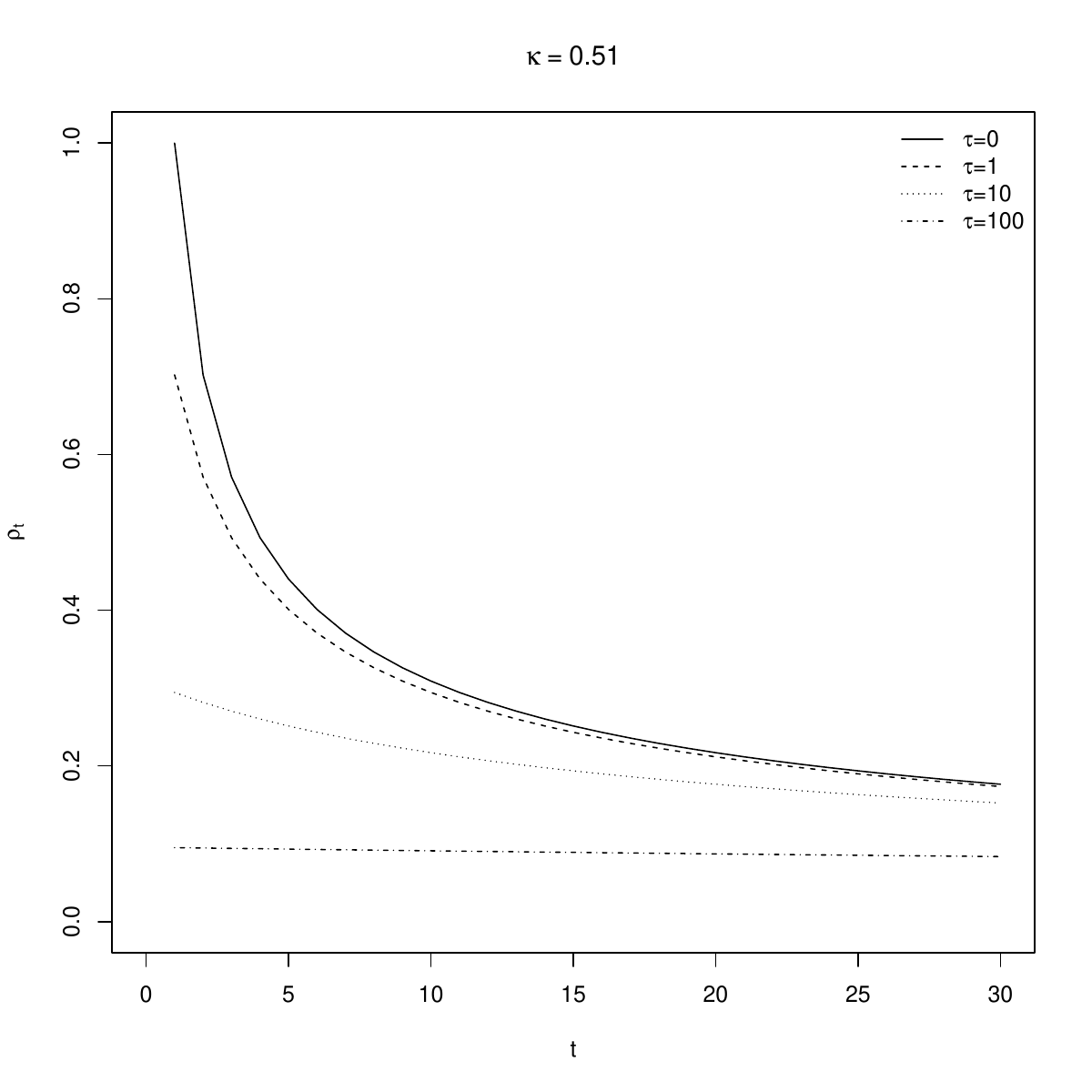} &
    \includegraphics[width=.4\textwidth]{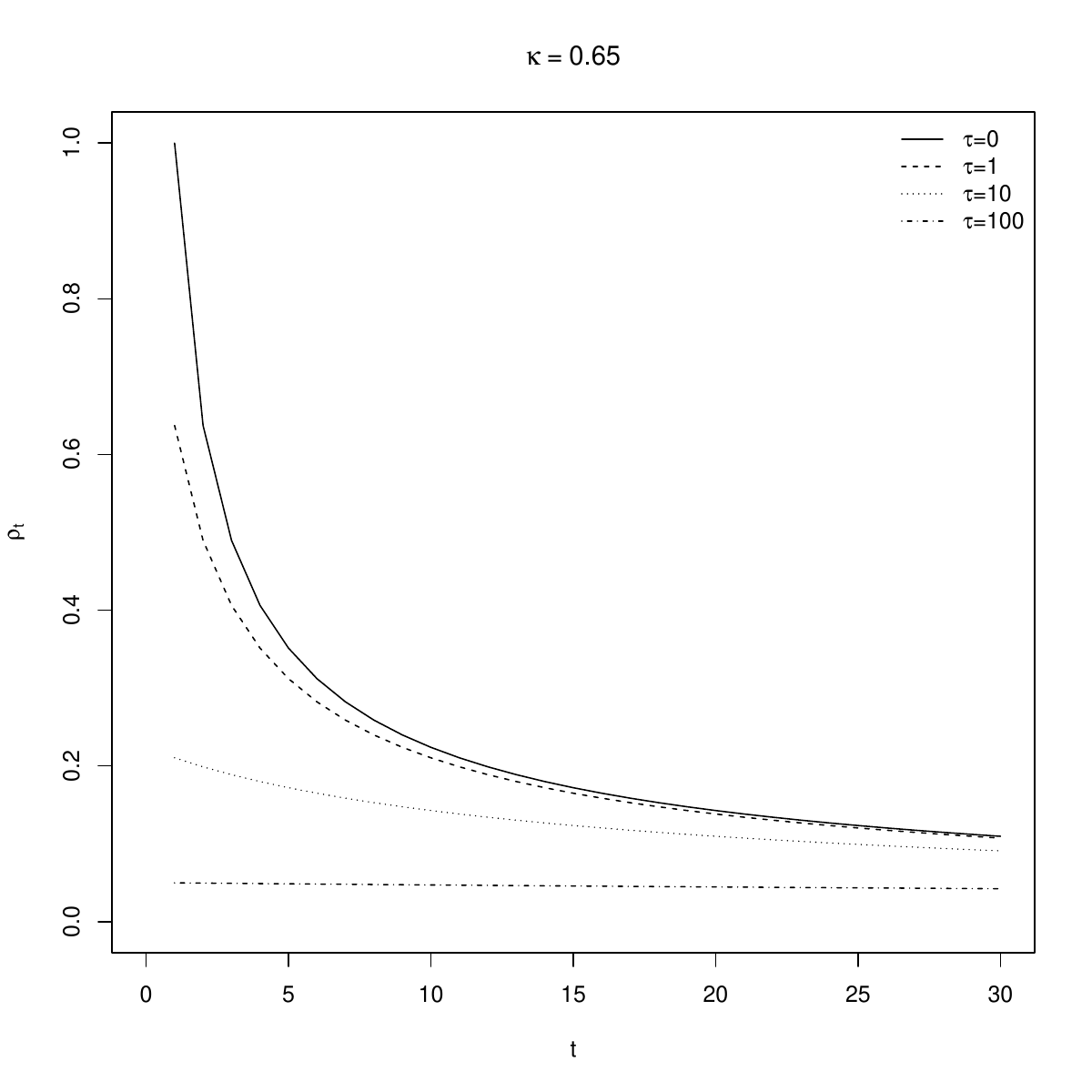}  \\  
    \includegraphics[width=.4\textwidth]{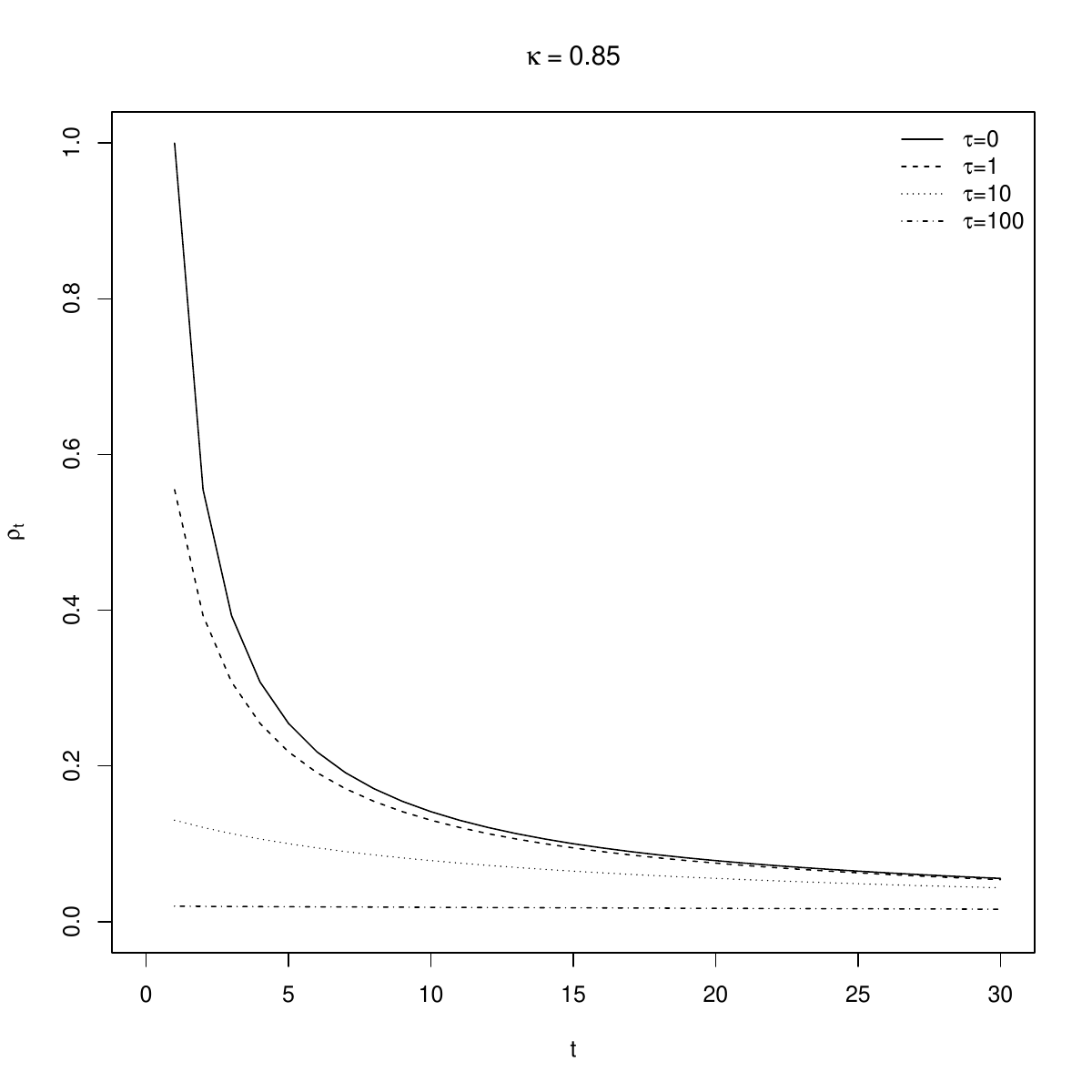} &
    \includegraphics[width=.4\textwidth]{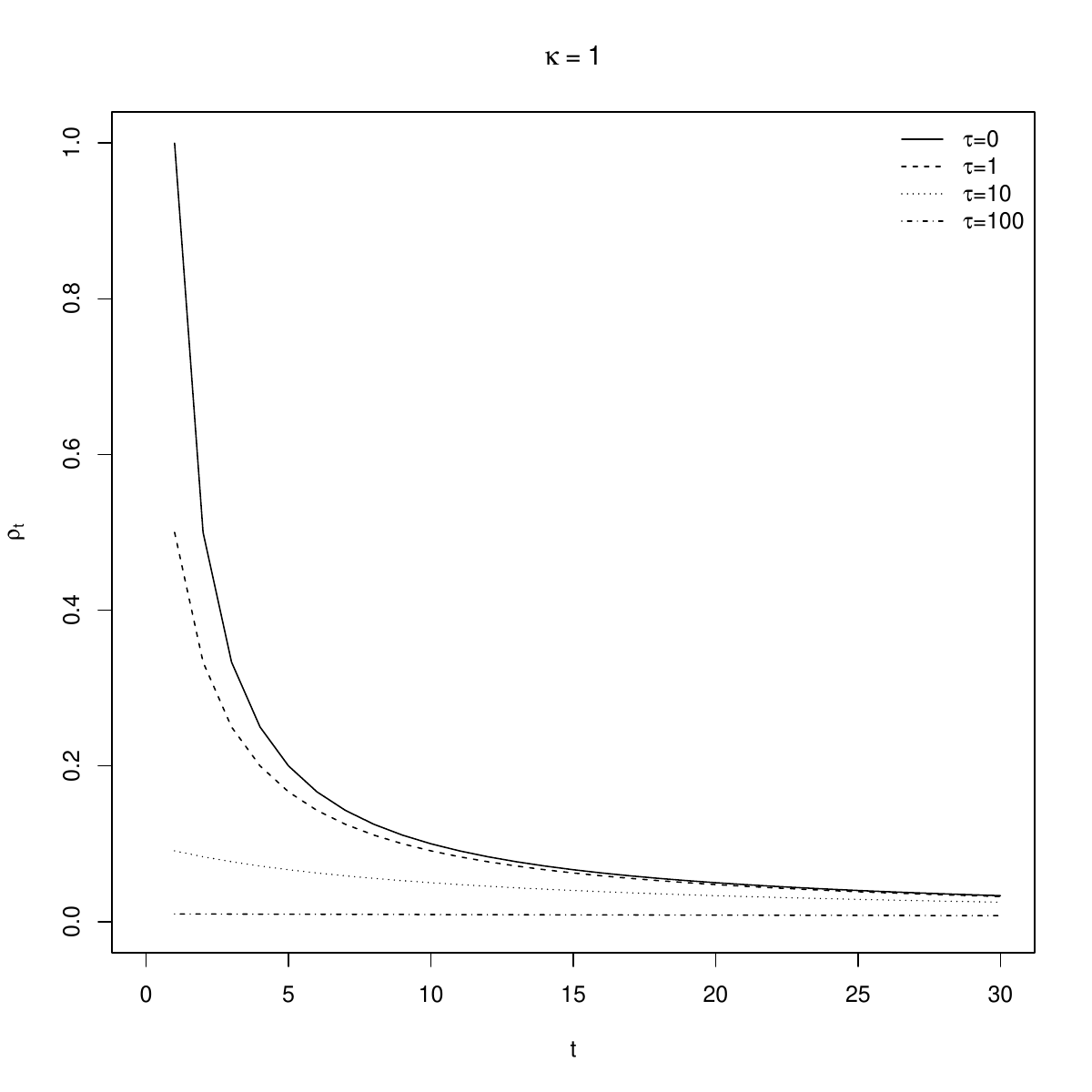}  
\end{tabular}
\caption{Evolution of the step size in the SGV algorithm under the scheme $\rho_t=(t+\tau)^{-\kappa}$ for different choices of $\kappa$ and $\tau$.}
\label{stepsizesfig}
\end{figure}

A common choice for the sequence of step sizes is given by $\rho_t=(t+\tau)^{-\kappa}$, where $\kappa\in\left(\frac{1}{2},1\right]$ represents the forgetting rate and $\tau\geq0$ is known as the delay. This scheme is illustrated in Figure \ref{stepsizesfig} for various choices of $\tau$ and $\kappa$.  From this figure is possible to observe that $\kappa$ determines how fast the step size declines, and $\tau$ affects the sequence starting level $\{\rho_t\}$.

Since the stochastic blockmodel satisfies the assumptions of the SGV algorithms, it is possible to derive the required updates from equations \eqref{vbgensolution} and \eqref{svbvupdate}.  One small caveat here is that, because of the matrix structure associated with the data, the mini-batches used to update the local parameters must contain more than one observation.  Different schemes for selecting the observations in the minibatch are possible, in this paper we adopt the simplest possible one, which corresponds to mini-batches formed of all observations involving a small subset of the nodes in the network.  The resulting algorithm is summarized as follows:
\begin{enumerate}[label*=\arabic*.]
\item Randomly initialize the global parameters $a^{\star (0)}_{k,l}$, $b^{\star (0)}_{k,l}$ for $1\leq k\leq l\leq K.$   
\item Repeat:
\begin{enumerate} [label*=\arabic*.]
\item Randomly obtain a subnetwork $S$ by sampling uniformly nodes in the original network. 
\item Update the local variational probabilities $q(\xi_i=k)$, for all $i\in S$ and all $k$, using equation \eqref{localvbupdate} with the corresponding scaling factor. 
\item Compute the intermediate global parameters $\hat{a}^\star_{k,l}$, $\hat{b}^\star_{k,l}$ using the noisy gradient in equation \eqref{globalvbupdates}. 
\item Update the estimates of the global variational parameters using \eqref{svbvupdate}:
$$
a^{\star (t)}_{k,l}=(1-\rho_t)a^{\star (t-1)}_{k,l}+\rho_t\mathcal{C}\hat{a}^\star_{k,l}\,\,\, \text{and}\,\,\, b^{\star (t)}_{k,l}=(1-\rho_t)b^{\star (t-1)}_{k,l}+\rho_t\mathcal{C}\hat{b}^\star_{k,l}.
$$
\end{enumerate}
\end{enumerate}

In terms of the stopping rule for the algorithm, notice that some of the computations required to calculate the ELBO are precisely those avoided by the SGV algorithm. For this reason, in order to asses convergence of the algorithm, we track a noisy estimate of the ELBO computed over a fixed subnetwork composed of randomly selected nodes.  We generate the subnetworks used in a sweep of the stochastic gradient algorithm by first partitioning the nodes into $\lceil I/\omega \rceil$ blocks of size $|S| \approx \omega I$.  Once all $\lceil I/\omega \rceil$ sweeps have been completed (an \emph{epoch} in the usual stochastic gradient nomenclature) we can guarantee that the algorithm has updated all local parameters (i.e., all component indicators).  Except when noted otherwise (e.g., when we use a time budget as the stopping rule for the algorithm), we set up the algorithm to complete a minimum of 3 epochs before checking for convergence of the algorithm.  After that, convergence is checked after each additional epoch is completed.

Finally, it is worthwhile mentioning that the multimodality of the problem makes both MCMC algorithms and variational approximation algorithms susceptible to initial conditions. This problem is particularly salient in the standard variational Bayes, and is slightly ameliorated with the noisy gradient estimates of stochastic variational Bayes. Nonetheless, we employ multiple runs of the SGV algorithm initialized at different, randomly chosen configurations in order to evaluate the algorithm.

\section{Evaluation}\label{ilustsl}

\subsection{Simulated data}\label{D7Mssubec}

\begin{figure}[!htb]
\centering
\includegraphics[width=.5\textwidth]{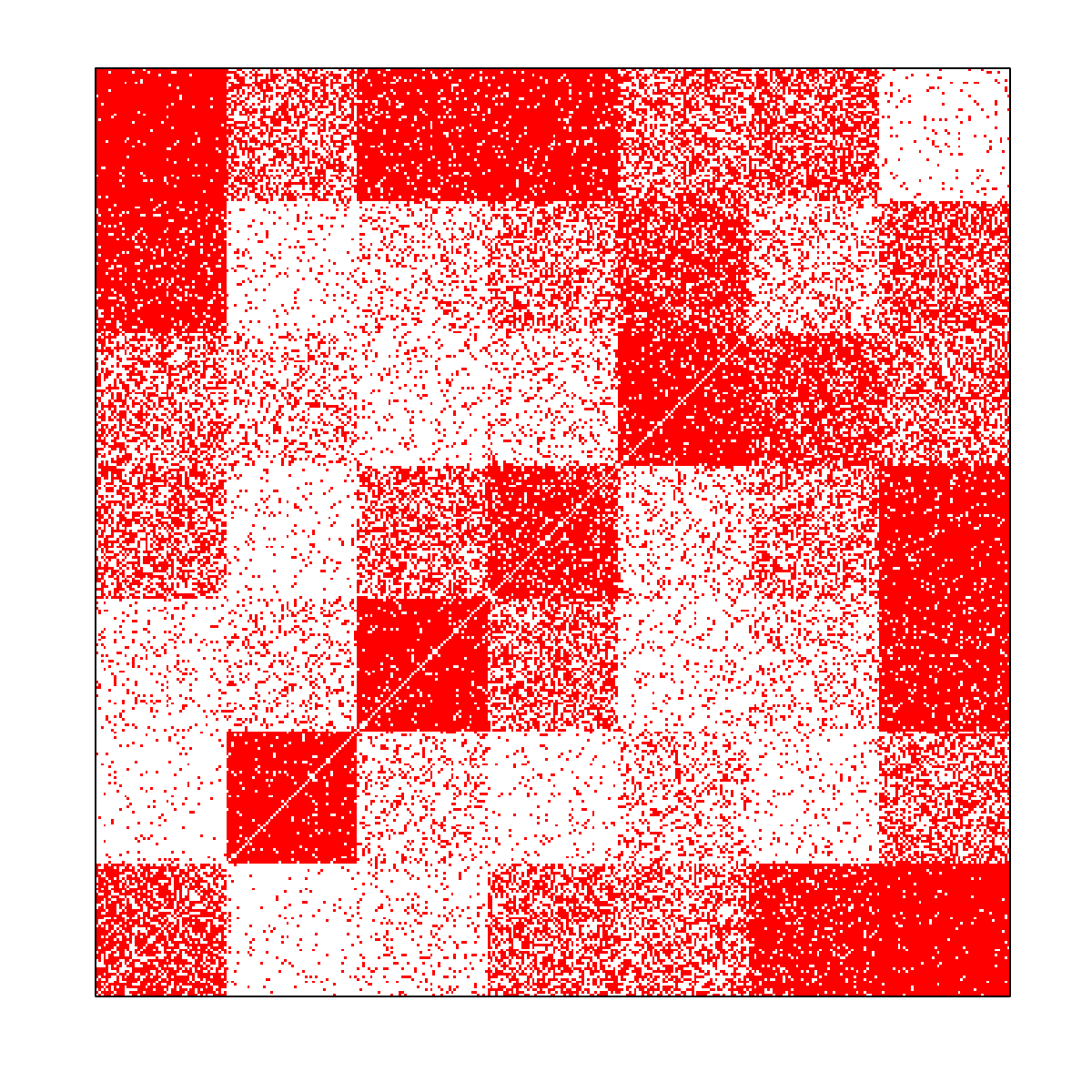}   
\caption{Pictorial representation of the adjacency matrix associated with the first simulated dataset. Actors in the network are placed along the $x$ and $y$ axis. $Y_{i,j}=1$ is represented by a red dot, while a lack of interaction is shown in white.}
\label{D7Mdata}
\end{figure}

In this section, we evaluate the performance of the SGV algorithm described in Section \ref{se:sgvb} in settings where the ground truth is known.  First, we make use of a simulated dataset which is represented graphically in Figure \ref{D7Mdata}. This network is constructed with $I=350$ actors, split evenly in $K^\star=7$ communities.  These communities are large and well-separated, so this is an ``easy'' case for community identification.  In the experiments throughout this section we set the maximum number of communities to $K=20$ and the model hyperparameters to $a=b=\alpha=1.$

\begin{figure}[!htb]
\centering
\includegraphics[width=1\textwidth]{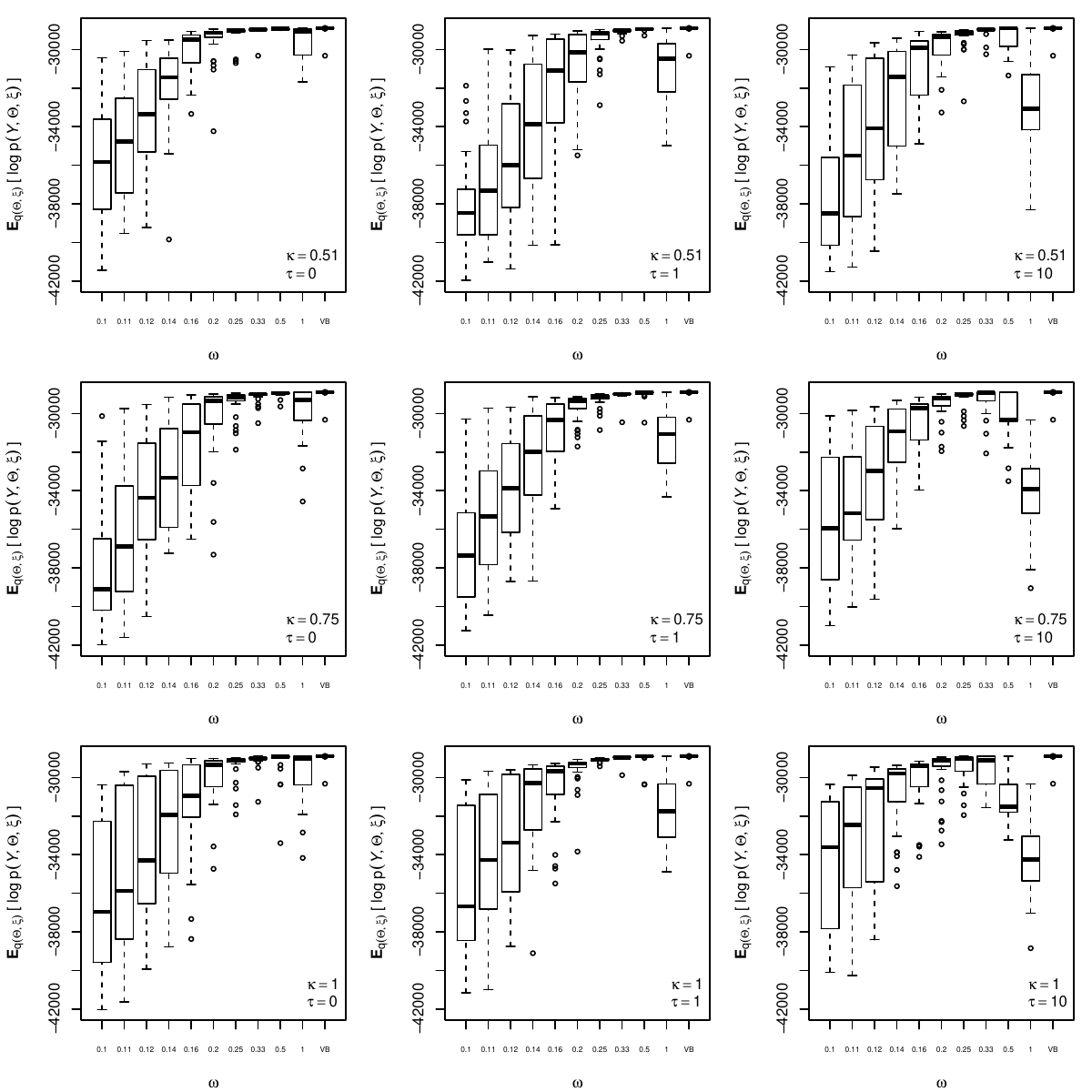} 
\caption{Boxplots summarizing the distribution of $F(q,\mathcal{Y})-H[q(\cdot)]$ for 32 initial conditions using the fist simulated dataset, with distinct parameter configurations and values of $\omega=|S|/I$. For every initial condition, the standard variational Bayes algorithm is executed until convergence. Then, the corresponding SGV algorithms are run for as much time as the variational algorithm.}
\label{D7Mcompprops}
\end{figure}

The first issue we address is the selection of the parameters in the sequence $\{\rho_t\}$ and the block size $|S|$.  In principle, any sequence satisfying conditions \eqref{conditionsrho} leads to an algorithm that is guaranteed to converge to a local mode.  However, the choice of these tuning parameters, as well as the batch size $\omega$, can have an important effect in the rate of convergence.  Using this simulated dataset and fixing the form of $\rho_t=(\tau+t)^{-\kappa}$, Figure \ref{D7Mcompprops} compares different choices of $\tau\geq0$ and $\kappa\in\left(\frac{1}{2},1\right]$.  For each combination of parameters, the algorithm is run a total of 32 times using different, randomly selected initial configurations.  The SGV algorithm is allowed to run for a \emph{time budget} equal to the (average) time that the standard variational algorithm takes to converge.  The free energy and entropy are then calculated using the complete network to provide a fair and consistent assessment of algorithmic performance.  As an additional reference, we also include the results associated with the regular variational algorithm.

A number of insights arise from these plots, which involve almost 3,000 runs of the algorithm. First, note that there is a huge variability in the quality of the solution generated by different initial configurations, no matter what parameter settings are used.  Furthermore, and not surprisingly, that variability tends to increase as $\omega$ decreases.  Secondly, note that the case of $\omega=1$ does not correspond to the variational Bayes algorithm as, even though the gradient is calculated with the whole dataset, the algorithm takes only a partial step in the direction of the gradient.  Therefore, increasing the batch size provides diminishing returns.  Finally, the graphs suggest that, while the performance of the algorithm is relatively robust to the choice of $\tau$ and $\kappa$, small values of $\omega$ tend to provide poor results.  Indeed, for most values of $\tau$ and $\kappa$, selecting batches that contain 25\% to 33\% of the observations seems to lead to results that are comparable with those generated by the standard variational algorithm, but lower batch sizes lead to very poor approximations for the same amount of computing time invested.  This should raise concerns around current standard practices in the use of SGV algorithms, which typically rely on much small batch sizes.

\begin{figure}[!htb]
\centering
\includegraphics[width=.9\textwidth]{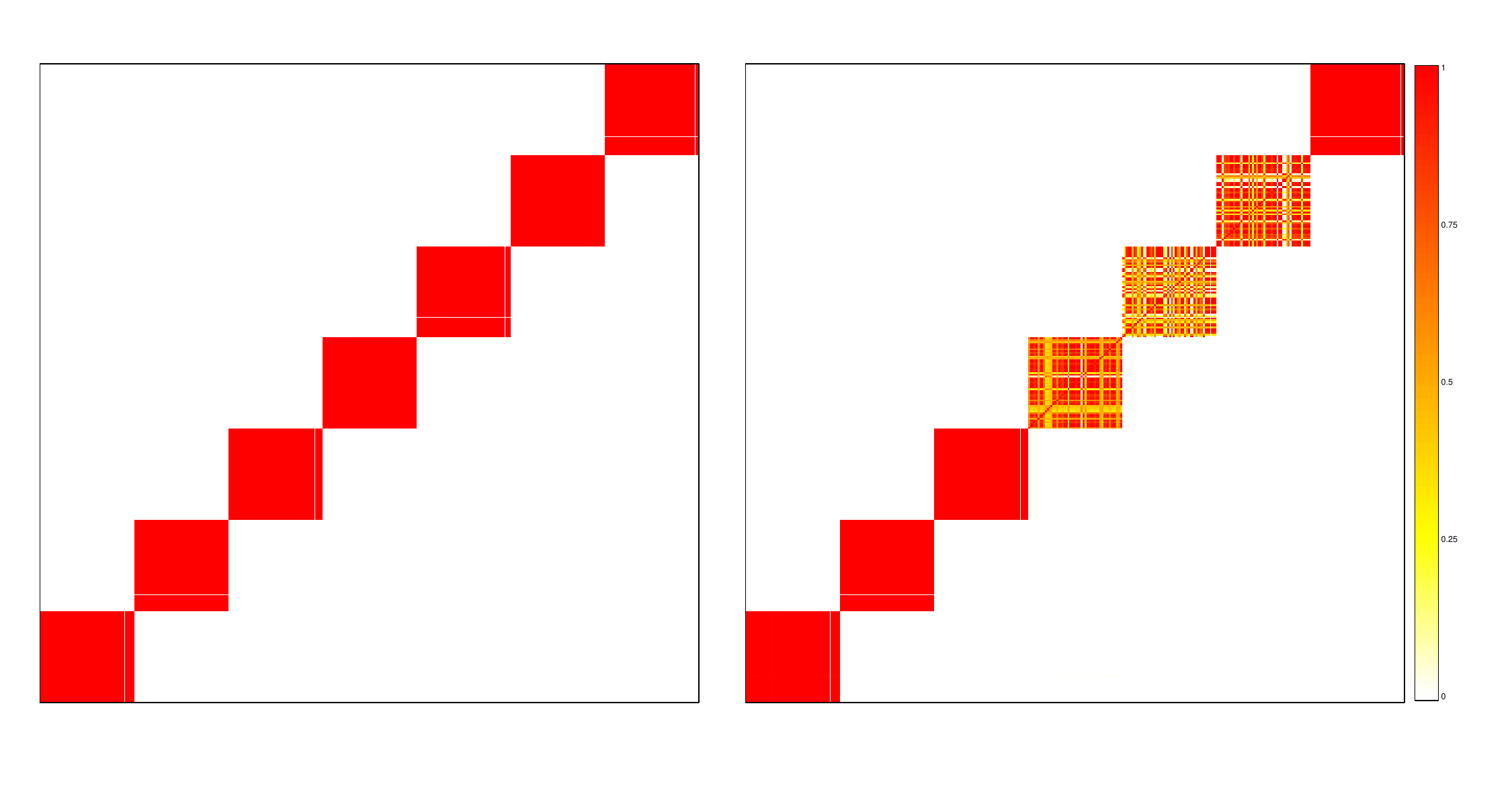} 
\caption{(Left) Monte Carlo estimates of pairwise co-clustering probabilities $\Pr(\xi_{i}=\xi_{j}\mid\mathcal{Y})$: $\text{ARI}=1$. (Right) Variational approximation $q(\xi_{i}=\xi_{j})$: $\text{ARI}=0.9$. }
\label{pwsD7}
\end{figure}

In this simulation, the second issue we investigate is the ability of the stochastic gradient version of the variational algorithm to recover the underlying community structure.  For this purpose we use the MCMC algorithm described in Section \ref{bb} (which we will treat as our gold standard) to obtain $100,000$ samples from the posterior distribution, and compare the resulting pairwise co-clustering probabilities $\Pr(\xi_i=\xi_j\mid\mathcal{Y})$ with those obtained from the best of 32 runs of the SGV algorithm with $\kappa=0.6$, $\tau=1$, and $\omega=0.25$ (see Figure \ref{pwsD7}).  It is clear that the MCMC algorithm is capable of fully recovering the underlying community structure in the network.  On the other hand, the SGV algorithm, although generating relatively good results, struggles to accurately reconstruct the clustering structure associated with 3 of the 7 communities.  Since in this simulated dataset the true vertex partition is known, we expand our comparison by first computing a point estimate of the partition structure under both algorithms using the approach described in \cite{Lau&Green07} with equal penalties for both misclassification errors, and then computing the adjusted Rand index \citep{Hubert&Arabie85} between each of those two partitions and the ground truth.  The ARI is a chance-corrected measure of similarity between two clusters constructed on the basis of pairwise agreements. An ARI of zero indicates that there is no more agreement than that expected by chance, while an ARI of one signifies that the two partitions are equivalent.  The ARI between the true vertex partition and the optimal partition under the MCMC algorithm is $1,$ while the ARI for the optimal partition under the variational approximation is $0.9$.  This loss in accuracy needs to be contrasted with the gains in computational speed.  The average run of a C implementation of the SGV algorithm for this dataset under the settings described above takes 15 seconds on a standard laptop with $8$GB of RAM and a $2.66$GHz Intel Core i7 processor.  In contrast, it takes approximately 4.5 hours for the MCMC under similar conditions.  Although this is not intended as a formal algorithm efficiency comparison, it does give a rough idea of the significant difference in computational time between the two approaches.

\begin{figure}[!htb]
\centering
\begin{tabular}{cc}
\includegraphics[width=.49\textwidth]{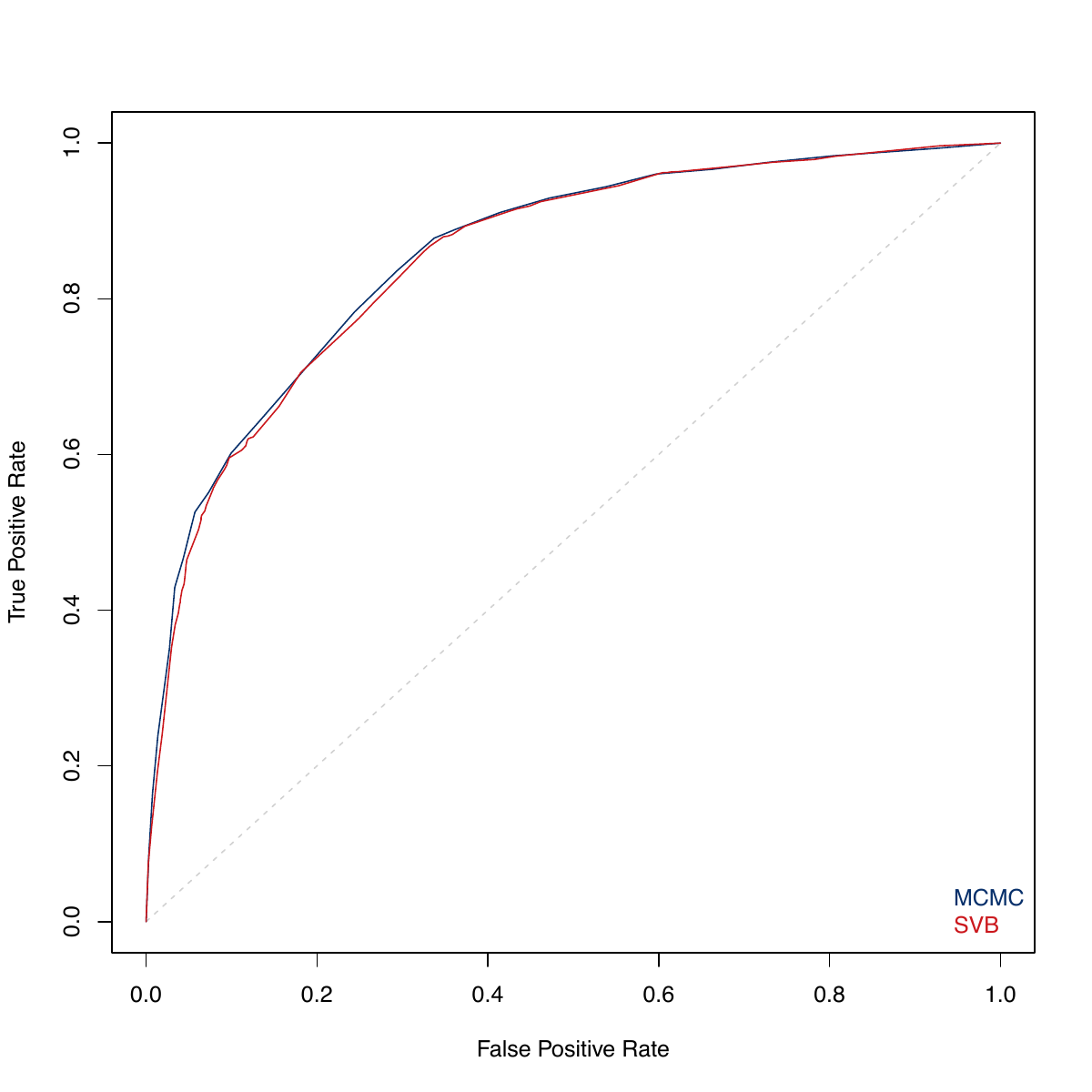} &
\includegraphics[width=.49\textwidth]{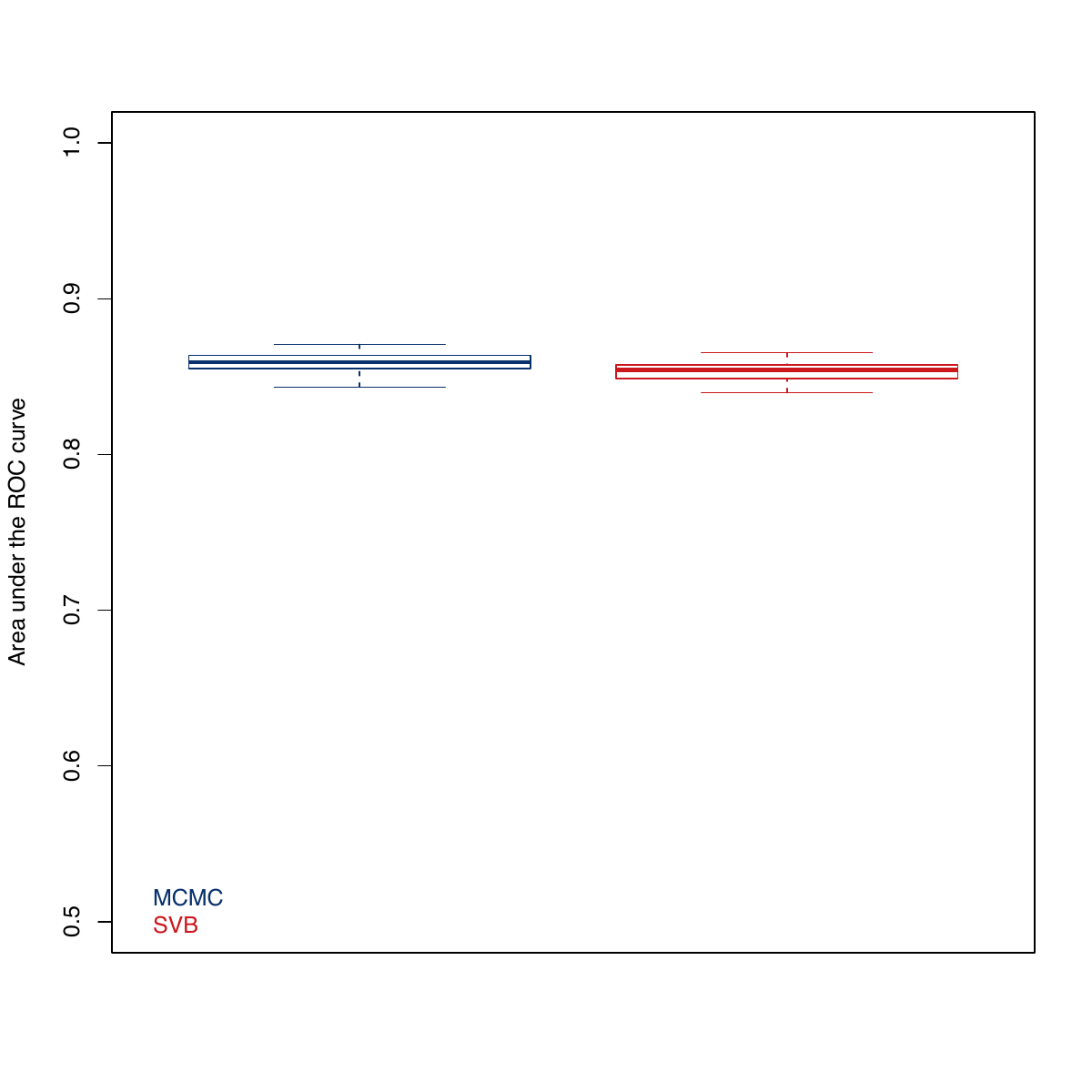}
\end{tabular}
\caption{(Left) Receiver operating characteristic curves for a typical validation subset. (Right) Boxplots of the area under the ROC curve for the MCMC and SGV algorithms.}
\label{D7MP}
\end{figure}

Next, we evaluate the predictive performance of the SGV algorithm. To this end, we carried out a twenty-fold cross validation exercise where, for each validation subset, we calculated the \emph{receiver operating characteristic} (ROC) curve, and its corresponding  \emph{area under the curve} (AUC).  As before, we compare the results obtained for the SGV algorithm against those generated by using the MCMC algorithm.  Figure \ref{D7MP} shows that, in this case, the SGV algorithm performs reasonably close to the MCMC in terms of prediction.  Furthermore, it is worthwhile mentioning that even in the case where the true underlying community structure is fully recovered, the ROC curve is not necessarily that of the perfect classifier.  This is because in this dataset some of the community interaction probabilities are, in fact, close to 0.5.

\begin{figure}[!htb]
\centering
\includegraphics[width=.5\textwidth]{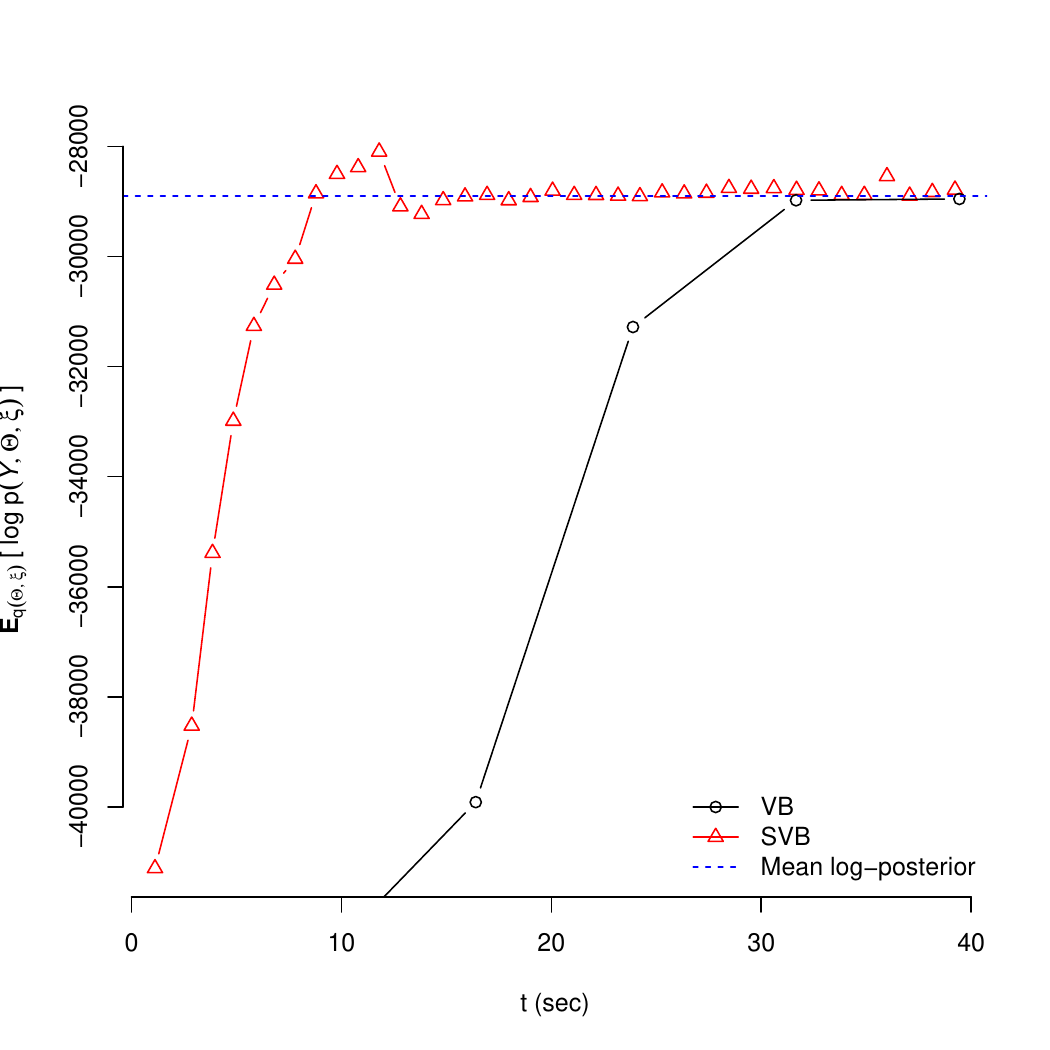} 
\caption{Evolution of $\mathbb{E}_{q(\Theta,\boldsymbol{\xi})}[\log P(\mathcal{Y},\Theta,\boldsymbol{\xi})]$ with respect to execution time in seconds. As before, $\omega=0.25,$ $\tau=.6$ and $\kappa=1$ in the SGV algorithm.}
\label{D7Mcompts}
\end{figure}

In order to better understand the performance of the SGV algorithm, we present in Figure \ref{D7Mcompts} a comparison of the evolution of the standard variational algorithm and its stochastic gradient version.  The graphs shows the value of the ELBO calculated over the complete network as a function of elapsed execution time.  The horizontal dashed line in this plots corresponds to the value obtained by averaging $\log p(\mathcal{Y},\Theta,\boldsymbol{\xi})$ evaluated over the posterior samples. This quantity, differs from $\mathbb{E}_{q(\Theta,\boldsymbol{\xi})}[\log p(\mathcal{Y},\Theta,\boldsymbol{\xi})]$, as the former averages over the posterior while the later takes the expectation with respect to the variational distribution.  However, these two quantities are in similar scales and, therefore, give an idea of how well the two variational algorithms approximates the posterior distribution.  From this figure it is possible to observe that, in this case, the stochastic variational optimization performs reasonably close to the MCMC algorithm, supporting the findings of Figures \ref{pwsD7} and \ref{D7MP}.  Furthermore, Figure \ref{D7Mcompts} suggests that, since the SGV algorithm climbs faster in the early stages, if the algorithm cannot be run until convergence, the stochastic version may be preferable.

\begin{figure}[!htb]
\centering
\includegraphics[width=.5\textwidth]{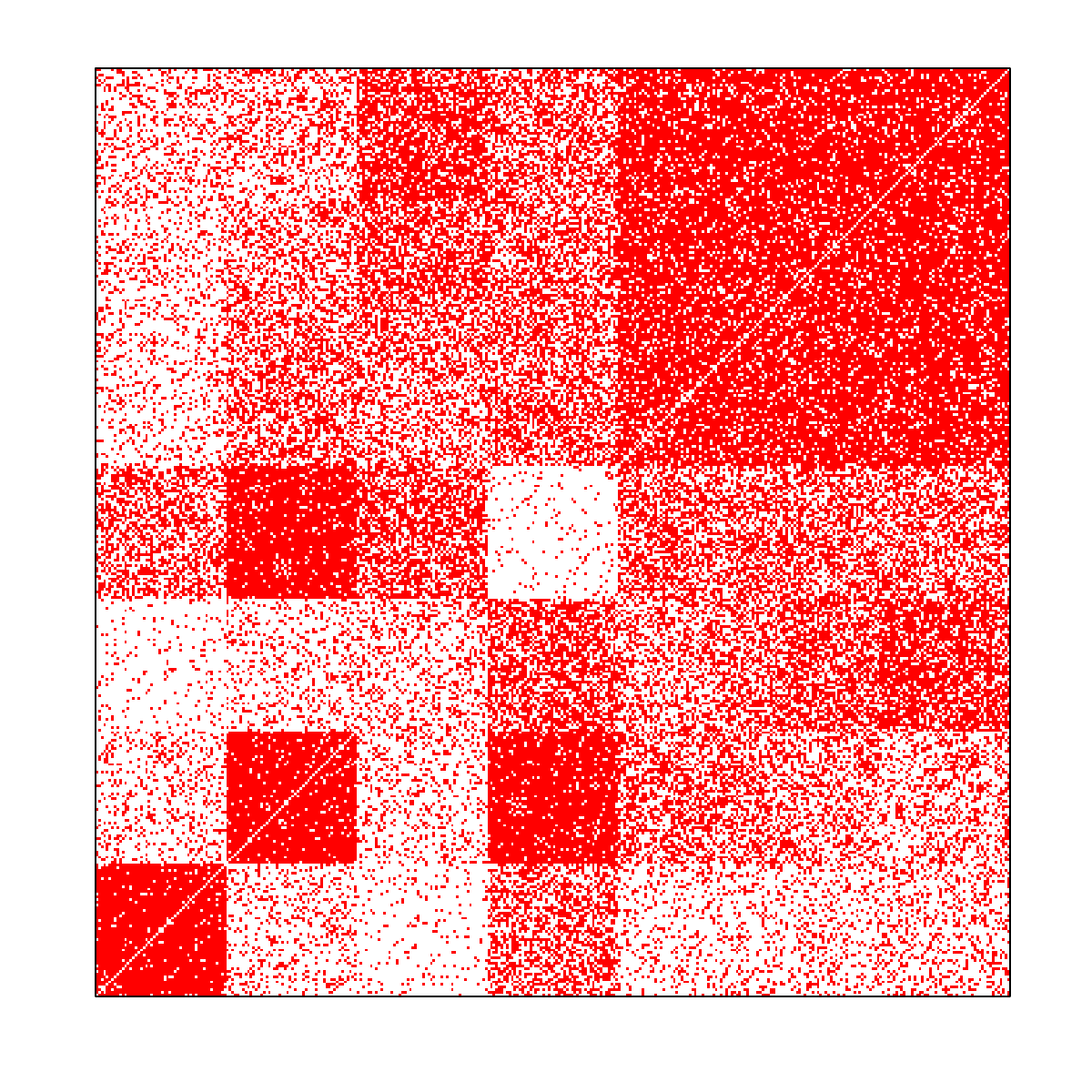}   
\caption{Adjacency matrix for the second simulated dataset. Actors in the network are placed along the $x$ and $y$ axis. $Y_{i,j}=1$ is represented by a red dot, while a lack of interaction is shown in white.}
\label{D7CC}
\end{figure}

For our second experiment, we repeat the previous exercise with another simulated dataset with a similar structure, i.e., $I=350$ individuals evenly split among $K^\star=7$ communities.  However, in this second dataset the communities are less clearly differentiated.  Specifically, we can see in Figure \ref{D7CC} that the interaction probabilities for the sixth community have been selected near the average of those corresponding to communities five and seven, making it hard to distinguish vertexes belonging to these three communities.

\begin{figure}[!htb]
\centering
\includegraphics[width=.9\textwidth]{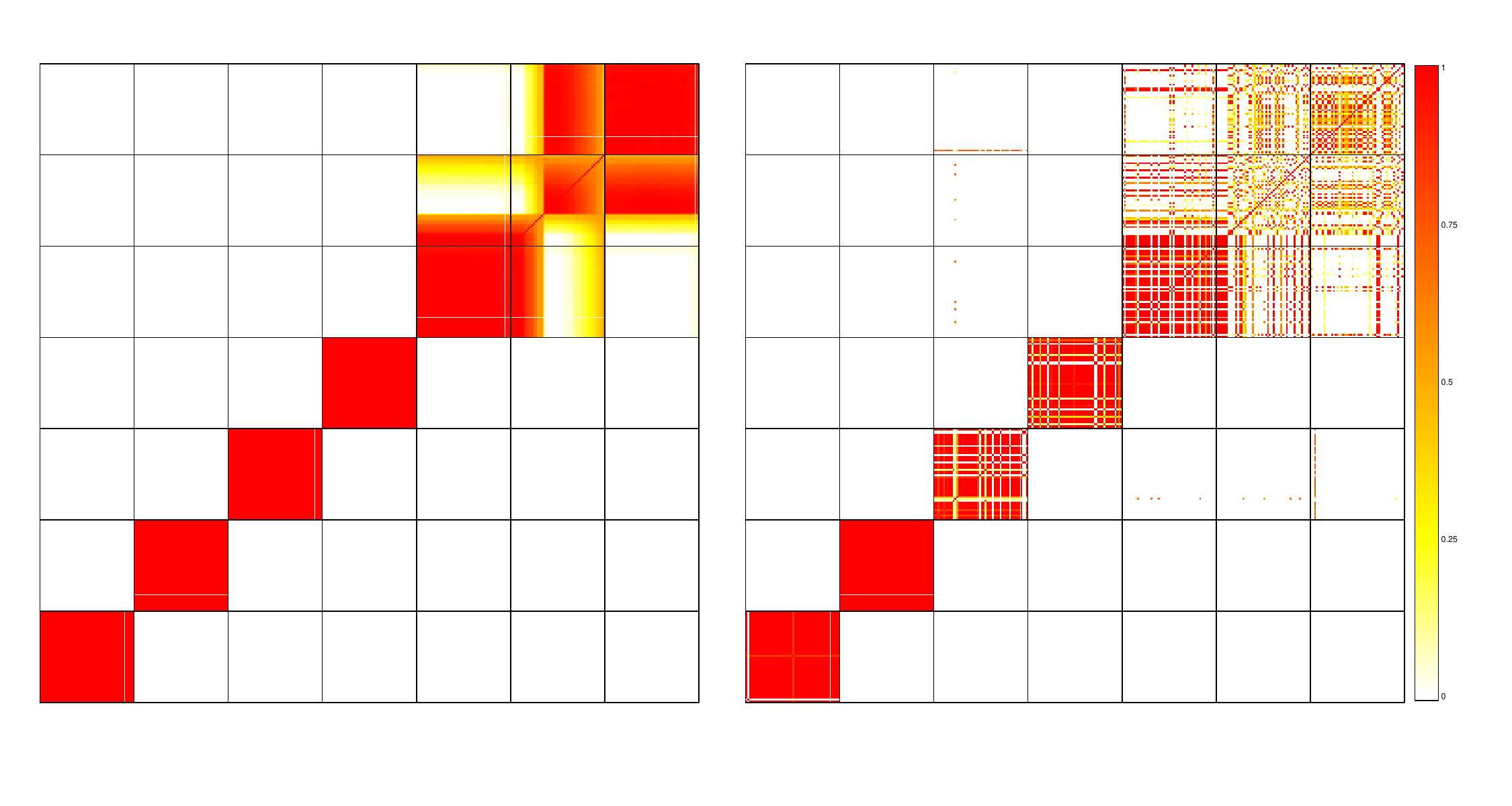} 
\caption{(Left) Monte Carlo estimates of co-clustering probabilities $\Pr(\xi_{i}=\xi_{j}\mid\mathcal{Y})$: $ARI=0.81$. (Right) Variational approximation $q(\xi_{i}=\xi_{j})$: $ARI=0.67$.}
\label{D7CCr}
\end{figure}

\begin{figure}[!htb]
\centering
\begin{tabular}{cc}
\includegraphics[width=.49\textwidth]{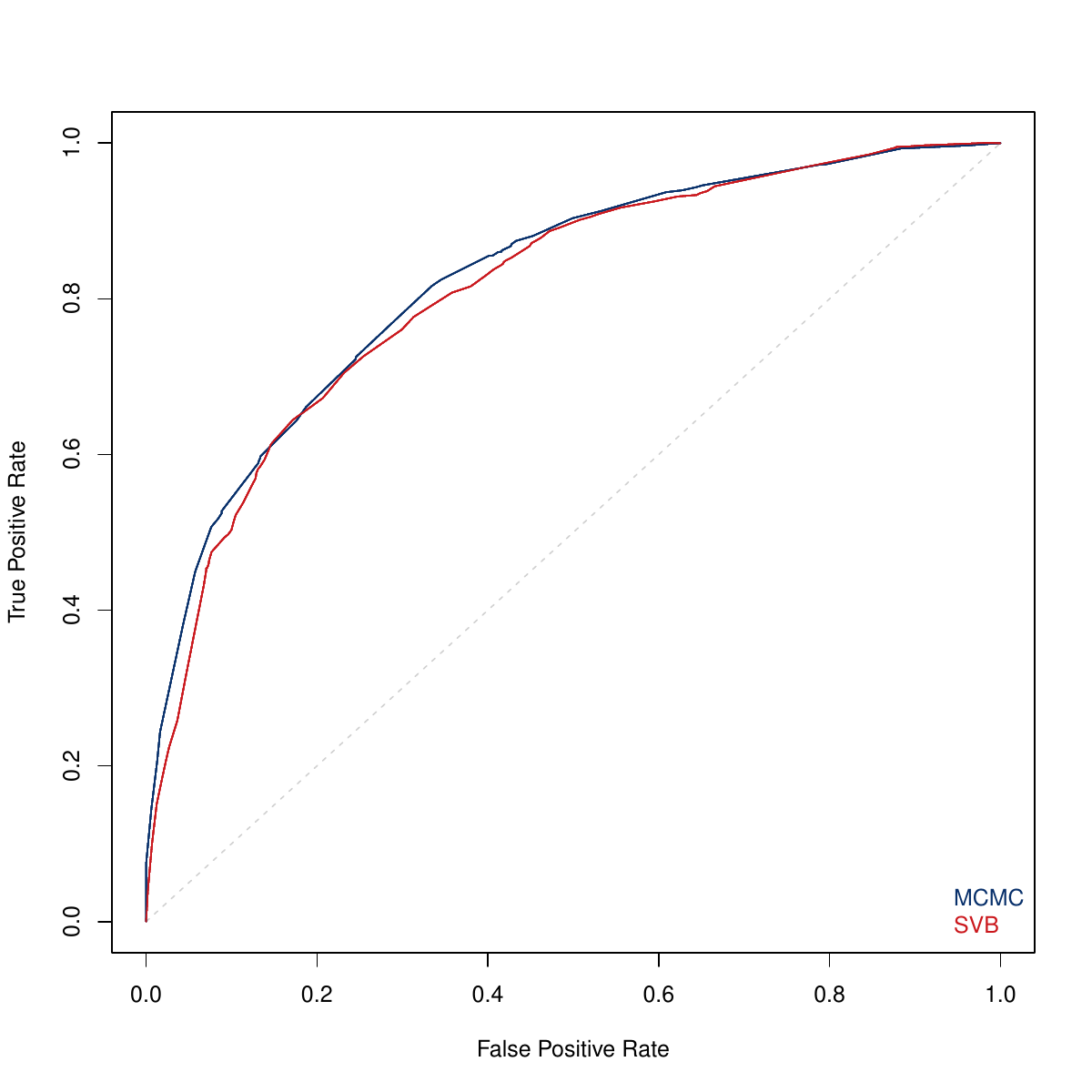} &
\includegraphics[width=.49\textwidth]{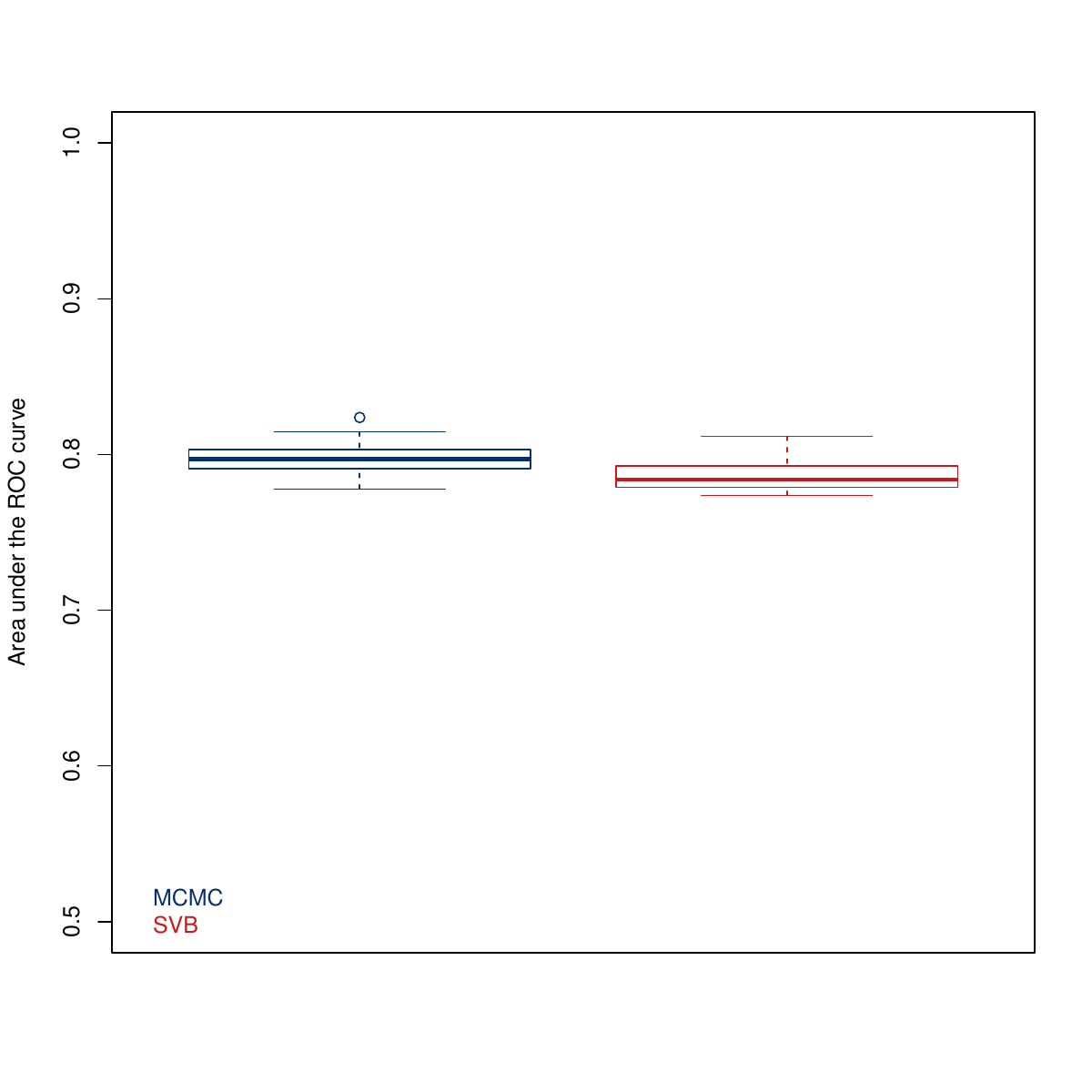}
\end{tabular}
\caption{(Left) Receiver operating characteristic curves for a typical validation subset. (Right) Boxplots of the area under the ROC curve for MCMC and SVB algorithms.}
\label{D7CCP}
\end{figure}

Figure \ref{D7CCr} shows the inferred community structures for this dataset under both computational approaches. In this case solid black lines are added to facilitate the recognition of the true community structure. From this figure, we see that that both algorithms have difficulties separating the individuals in the sixth community.  While the MCMC places part of these vertices in community five and the rest in community seven with high probability, the SGV algorithm tends to average over the modes mixing individuals from the three communities into a single cluster. The ARI between these point estimates to the inferred partition and the true community structure is $0.81$ and $0.67$ for the MCMC and SGV approximation, respectively.  In turn, from Figure \ref{D7CCP}, we see that the predictive performance is affected similarly for both algorithms. In both cases, there is a reduction in the AUC level, and a small increase in the variability is observed with respect to our first simulation.

The two simulation studies in this section suggest that, although the predictive performance of the SGV algorithms is similar to that of the MCMC algorithm when the number of communities is small and they are reasonably well separated, the ability of the variational algorithms to accurately identify communities is somewhat limited even in this relatively easy setting.

\subsection{Co-authorship network}\label{coanetvar}

\begin{figure}[!htb]
\centering
\includegraphics[width=.5\textwidth]{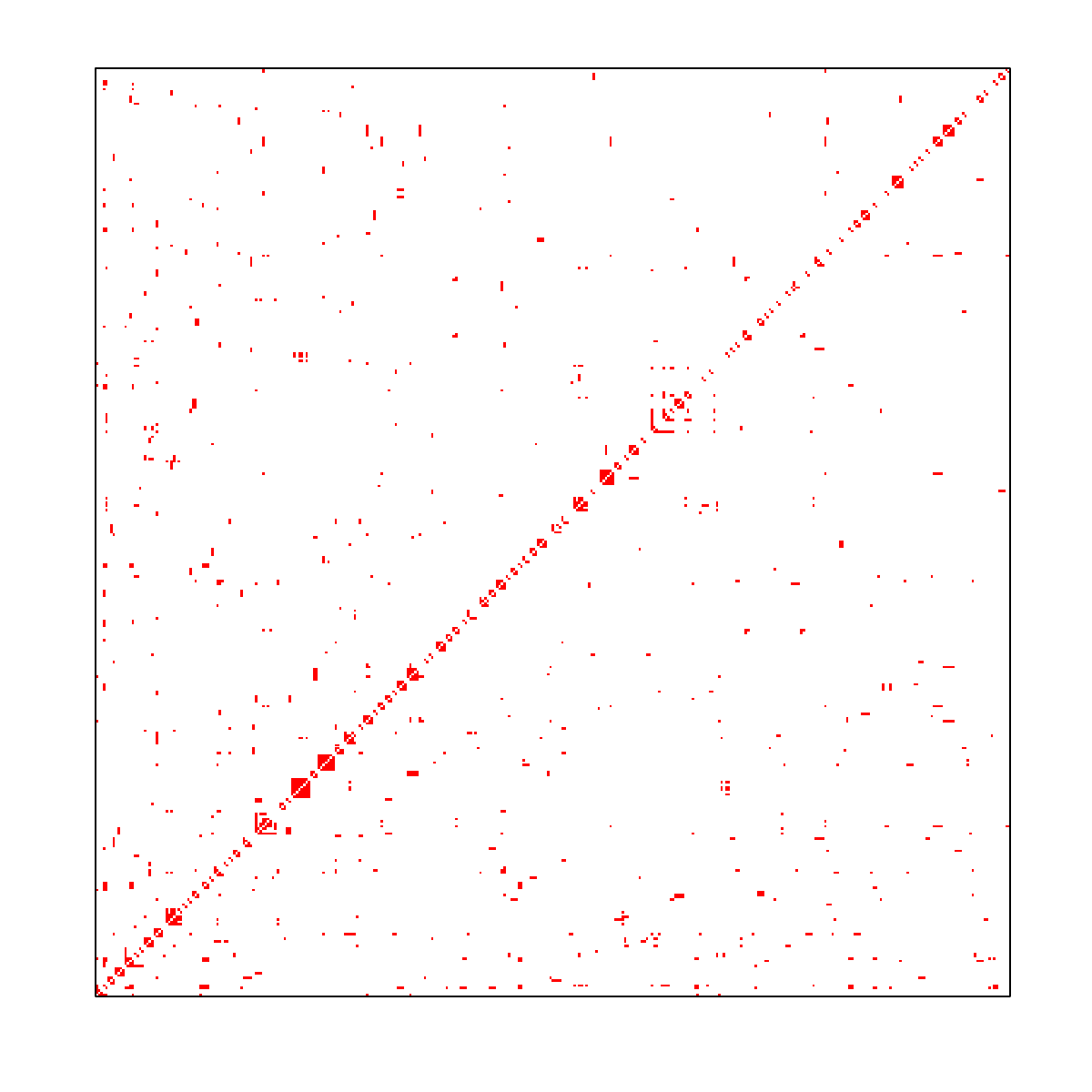} 
\caption{Raw data for the collaboration network of \cite{Newman06}. In this network vertices represent authors of scientific papers in the field of Network Science, and an edge represents the existence of at least one collaboration between those authors.}
\label{NSdata}
\end{figure}

In this Section, we consider the co-authorship network presented in \cite{Newman06}. In this network, vertices represent $(I=379)$ scientists working in the field of Network Science and an edge connecting two vertices indicates the existence of at least one co-authored paper that includes both scientists. This dataset is represented graphically in Figure \ref{NSdata} and includes a total of 914 publications up to early 2006, and is a subset of a larger network constructed from the bibliographies of two reviews, \cite{Newman03} and \cite{Boccalettietal06}.  For this dataset, we fit a stochastic blockmodel with  $K=30$ as the maximum number of communities and, as before, set the hyperparameters in the model are fixed to $\alpha=a=b=1$.

\begin{figure}[!htb]
\centering
  \begin{tabular}{@{}cc@{}}
    \includegraphics[width=.4\textwidth]{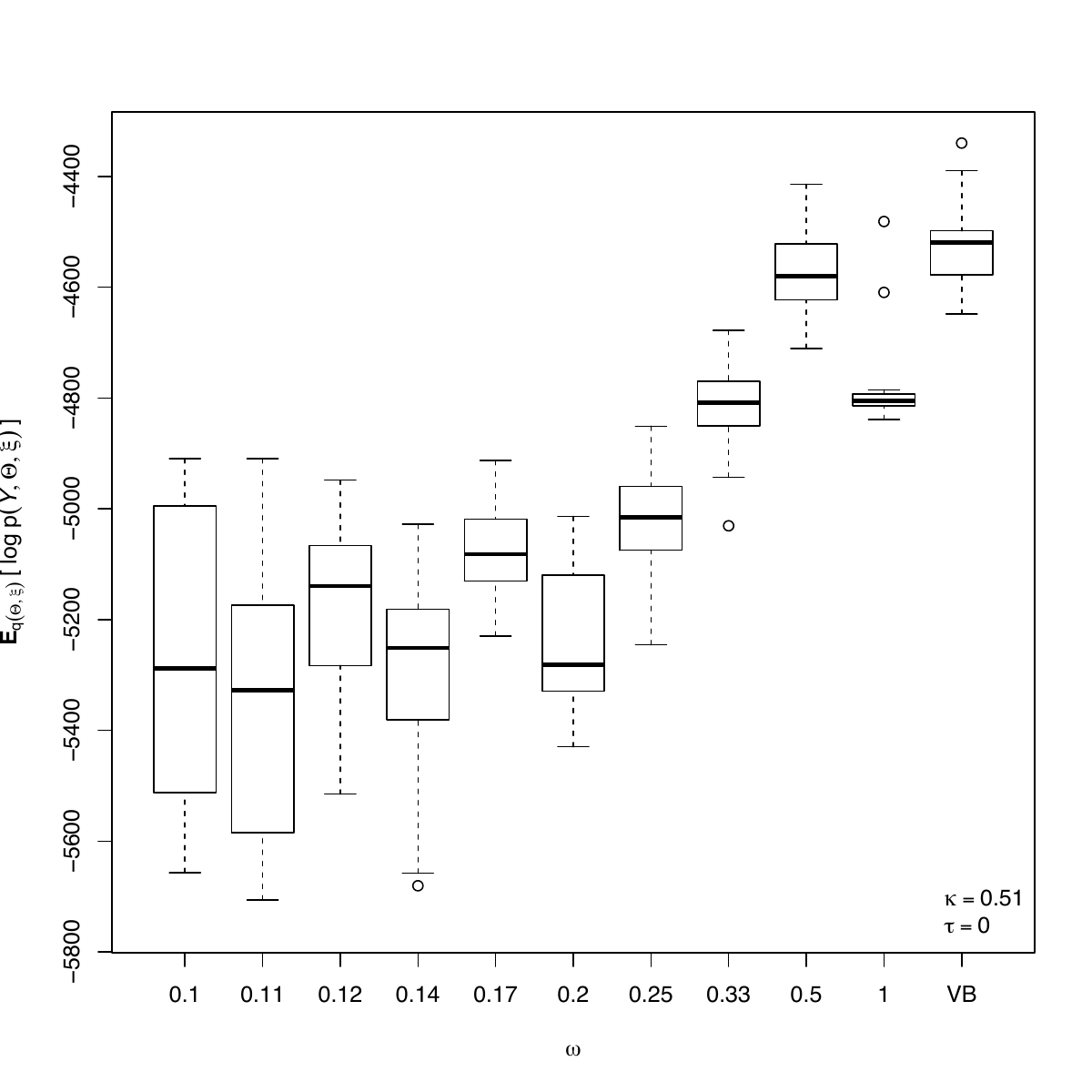} &
    \includegraphics[width=.4\textwidth]{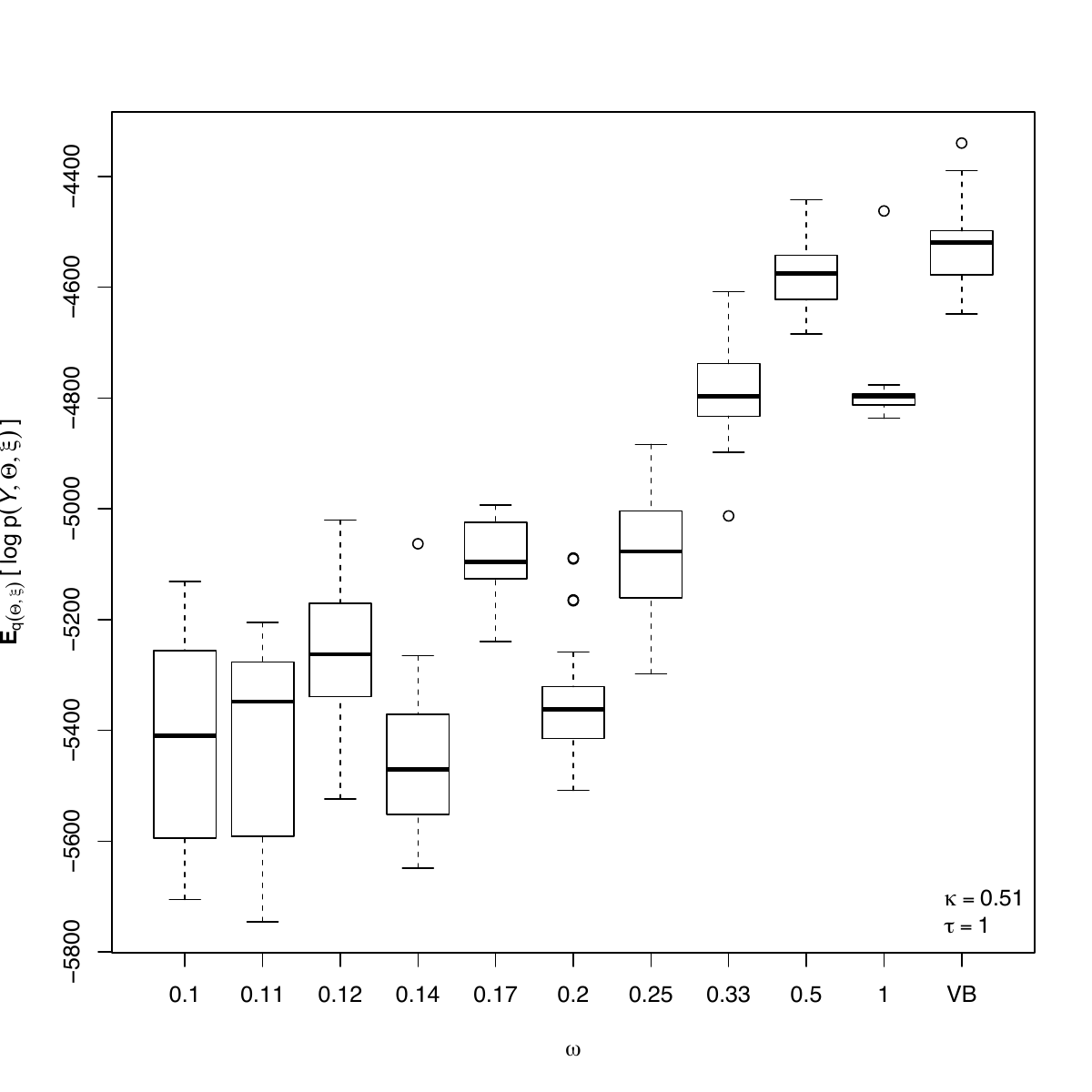} \\
    \includegraphics[width=.4\textwidth]{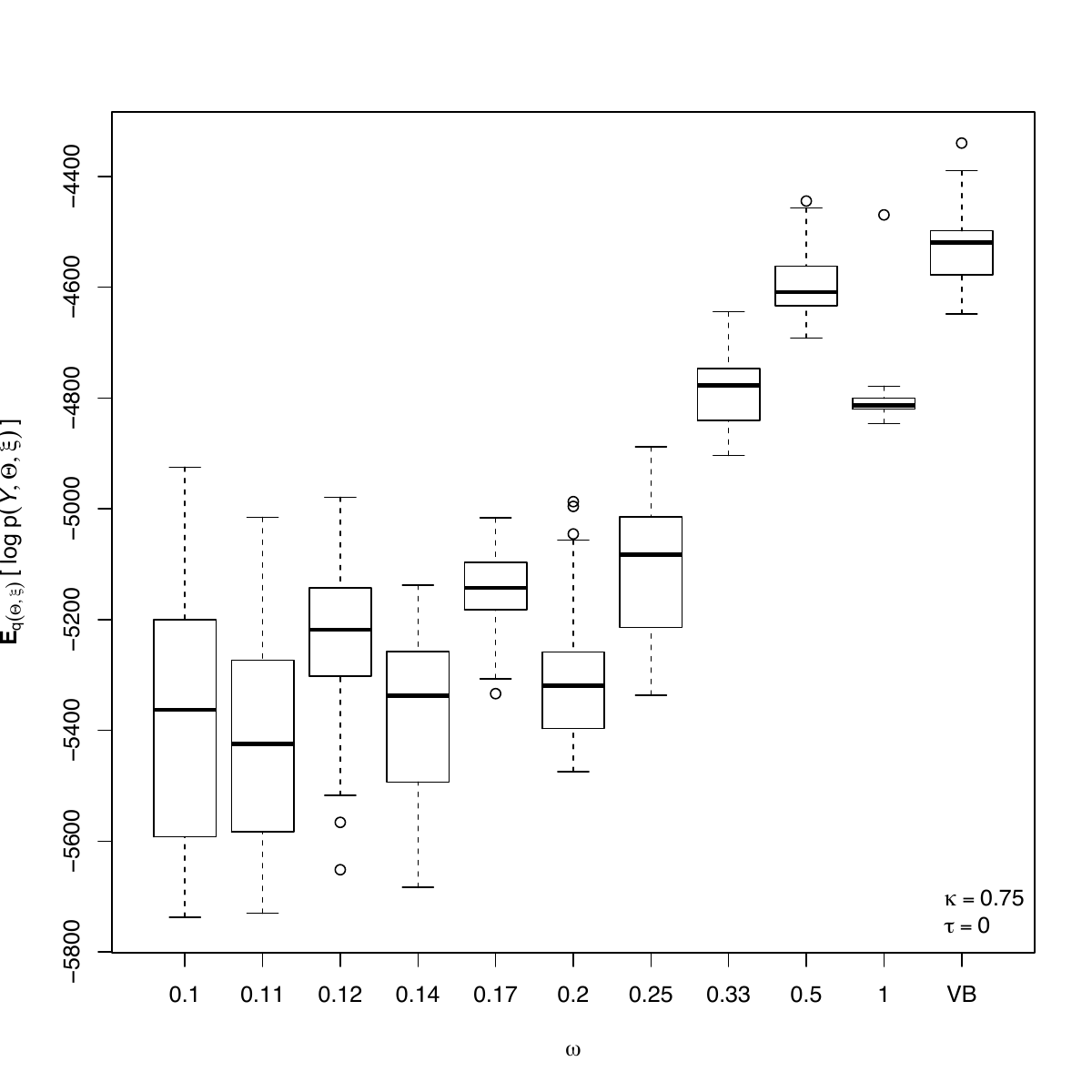}  &
    \includegraphics[width=.4\textwidth]{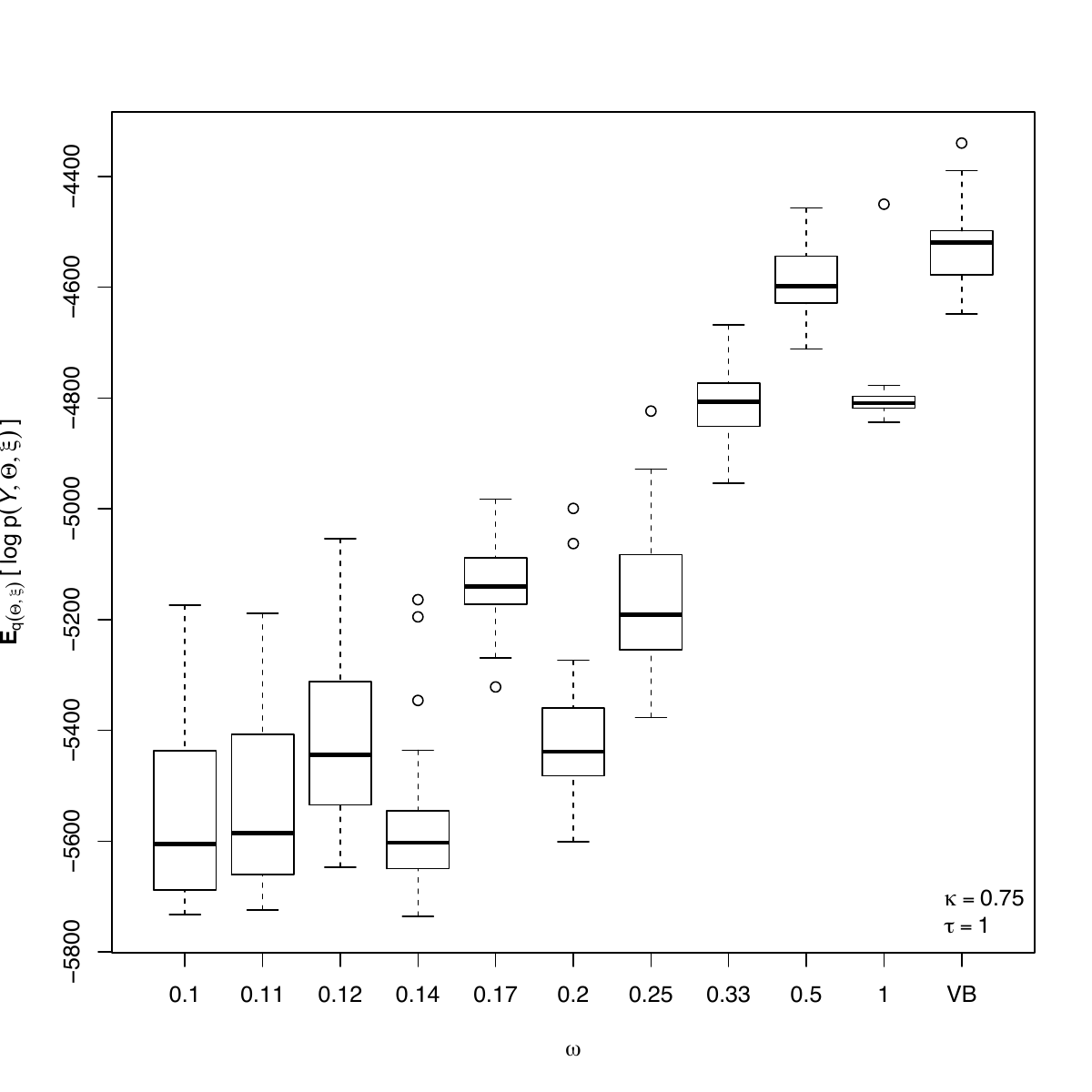} 
  \end{tabular}
  \caption{Boxplots summarizing the distribution of $F(q,\mathcal{Y})-H[q(\cdot)]$ for 32 initial conditions using the co-authorship network, with distinct parameter configurations and values of $\omega=|S|/I$. For every initial condition, the standard variational Bayes algorithm is executed until convergence. Then, the corresponding SGV algorithms are run for as much time as the variational algorithm.}
  \label{NScompprops}
\end{figure}

Figure \ref{NScompprops} shows the distribution of the ELBO for 32 randomly selected starting configurations and for different choices of the tuning parameters $\kappa$, $\rho$, and $\omega$.   As in the case of Figure \ref{D7Mcompprops}, all these runs use the same time-budget, which corresponds to the average run time of the regular variational algorithm.  Notice that, in this case, regardless of the choice of the step sizes, the size of the minibatch required to obtain results comparable to the non-stochastic variational algorithm is about half of the original network size.  This is consistent with our previous findings in the simulated datasets, and effectively limits the gains in computational efficiency derived from using a the stochastic gradient version of the algorithm.

\begin{figure}[!htbp]
\centering
\includegraphics[width=.75\textwidth]{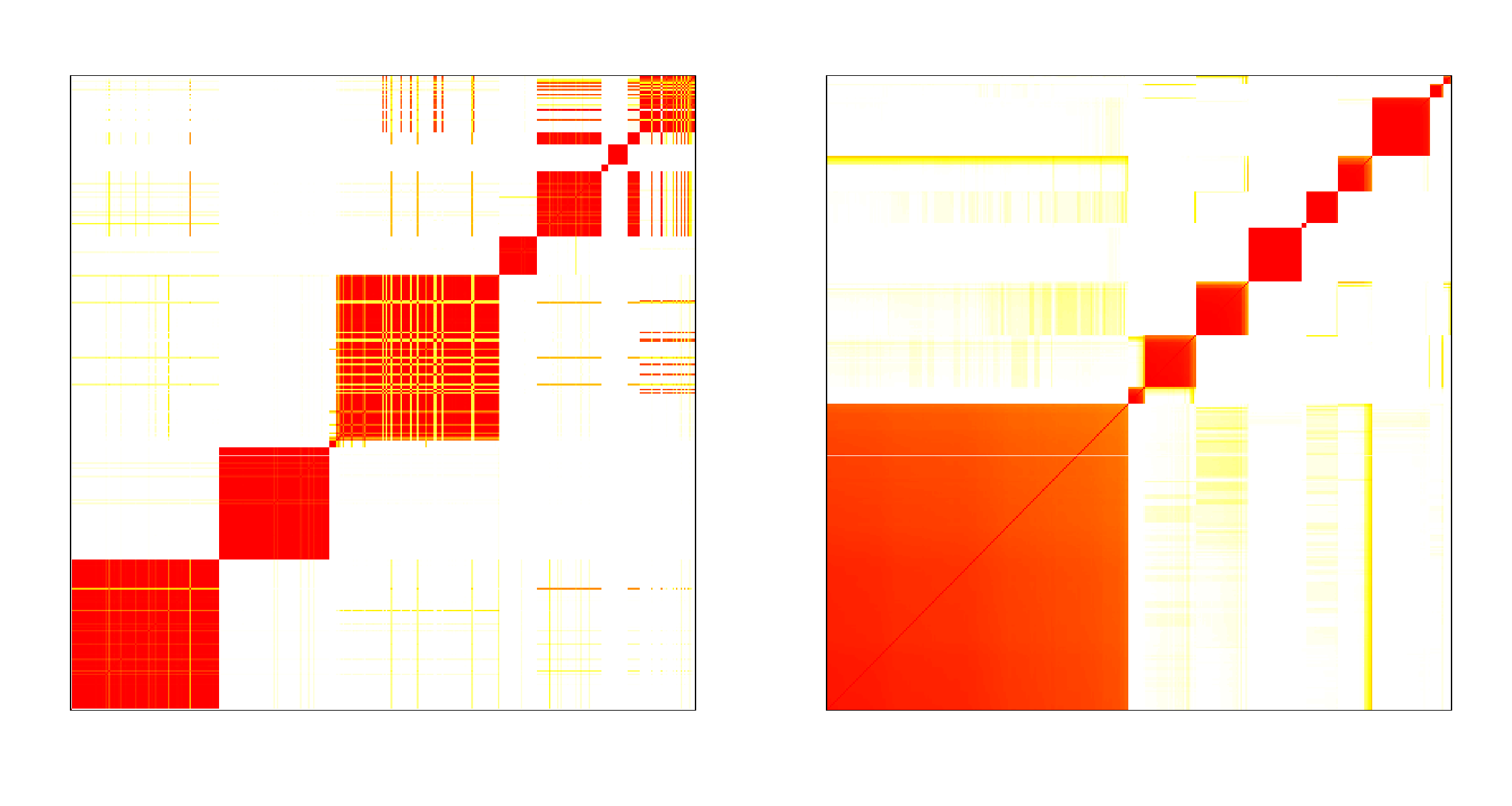} \\
\caption{Pairwise incidence matrices under MCMC (left) and stochastic variational approximation (right).} 
\label{NSresultsa}
\end{figure}

\begin{figure}[!htbp]
\centering
\includegraphics[width=.4\textwidth]{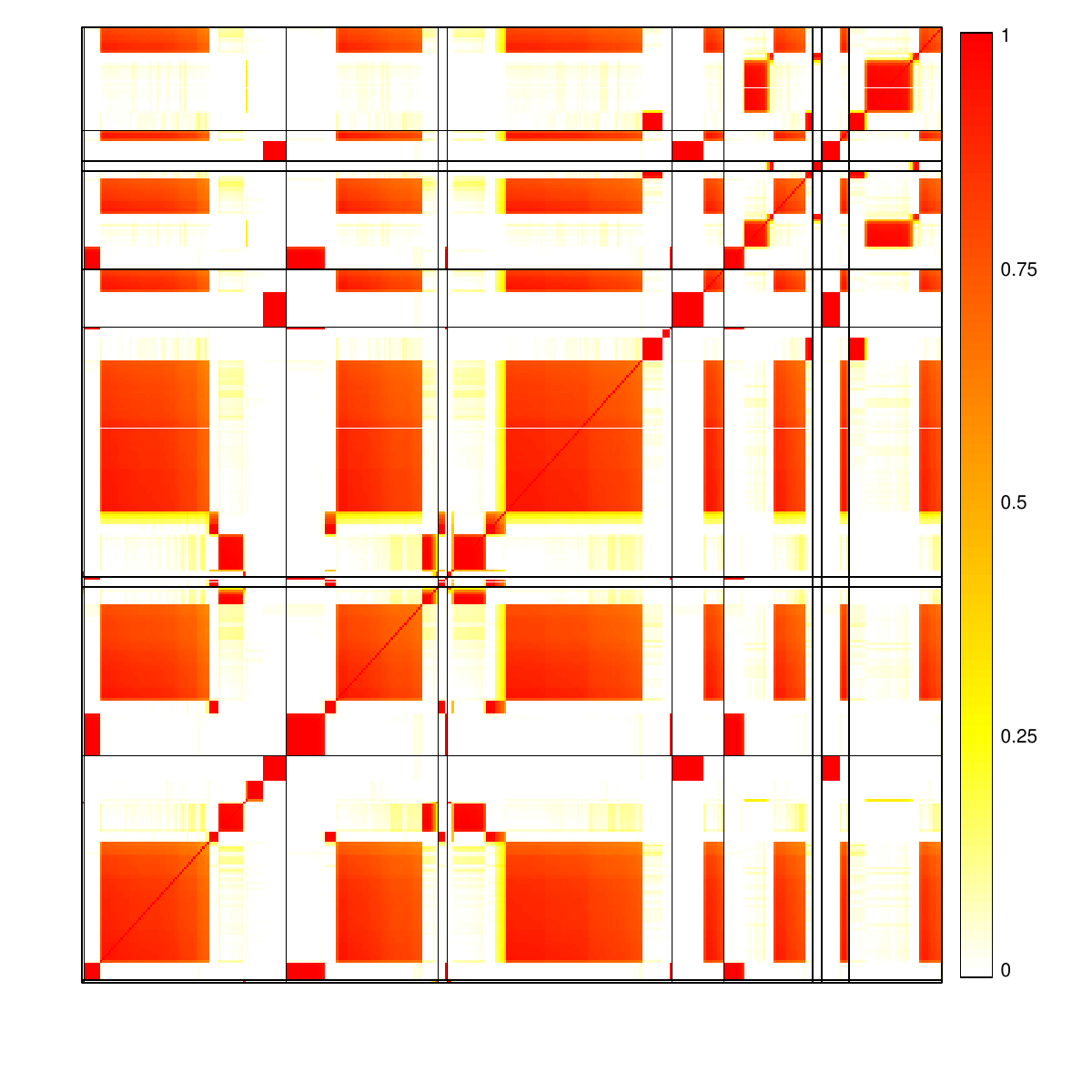} \\
\caption{Overlap in community structure from the two methods. This figure plot the incidence matrix from the SGV algorithm using the ordering from the MCMC.} 
\label{NSresultsb}
\end{figure}

Figure \ref{NSresultsa} shows the co-clustering matrices generated by the MCMC (based on the best of three runs, each consisting of $200,000$ iterations obtained after a burn-in period of $300,000$ samples) and the SGV algorithm (based on the best of 32 runs obtained with $\kappa=0.6$, $\tau=1$, and $\omega=0.25$).  Note that the ordering of the vertexes in both plots is different, and was selected in each case to make the corresponding graph as readable as possible.  To facilitate the comparison, Figure \ref{NSresultsb} shows the co-clustering matrix for the SGV algorithm (the same results as in the right panel of Figure  \ref{NSresultsa}), but with vertexes ordered as in the left panel of Figure \ref{NSresultsa}.  In contrast to the simulated datasets, in this example there is not a clear correspondence between the communities found by the two methods.  In particular, note that  white spaces in the anti-diagonal blocks correspond to MCMC communities being separated by the SGV approximation, while red areas outside of the main anti diagonal blocks indicate grouping of individuals in distinct MCMC communities.

\begin{figure}[!hb]
\centering
\begin{tabular}{cc}
\includegraphics[width=.49\textwidth]{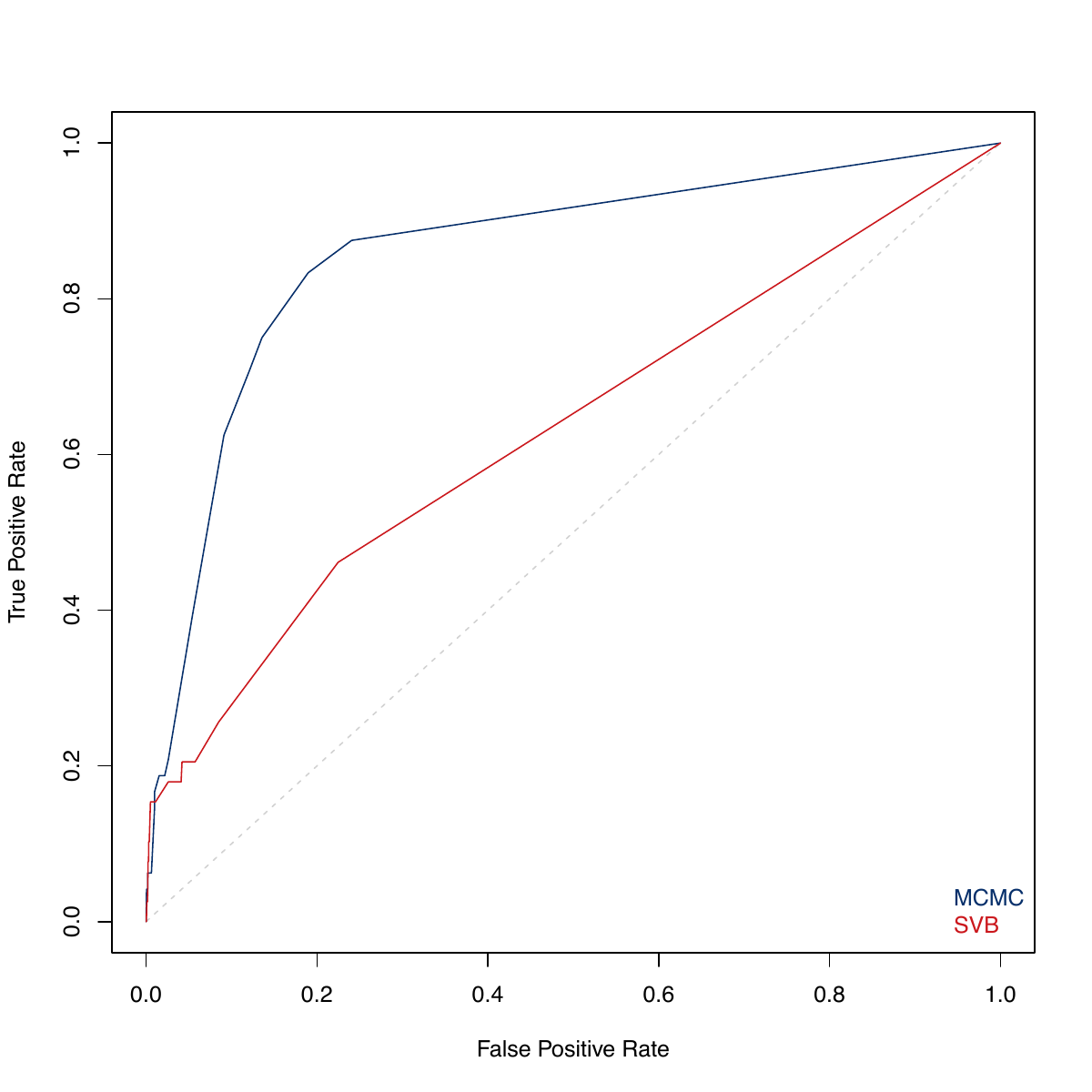} &
\includegraphics[width=.49\textwidth]{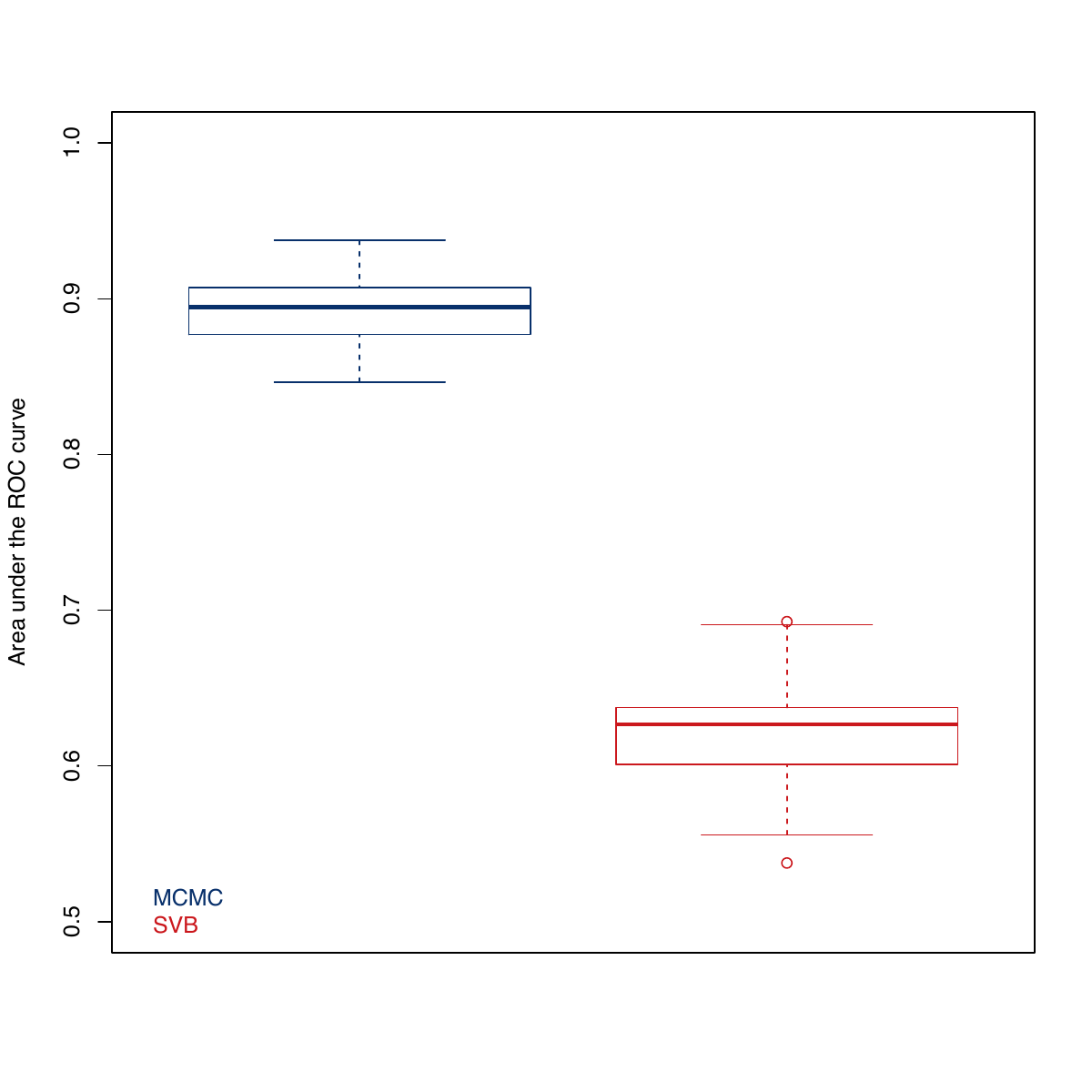}
\end{tabular}
\caption{(Left) Receiver operating characteristic curves for a typical validation subset. (Right) Boxplots of the area under the ROC curve for MCMC and SVB algorithms.}
\label{NSP}
\end{figure}

Since a true partition of the vertices is not available, we compare the quality of the partitions identified by the two algorithms in terms of their out-of-sample predictive performance.  Randomly selected subsets consisting of $5\%$ of the potentially observed links are treated as missing values and predicted in the same way as in Section \ref{D7Mssubec}. Figure \ref{NSP} shows the ROC curve for one of such subsets under both algorithms as well as boxplots corresponding to the area under the ROC curves for twenty of such test sets. From this figure it is clear that, for this dataset, the performance of the SGV algorithm is clearly suboptimal.

\subsection{IMDb network}\label{imdbsubsec}

\begin{figure}[!htb] 
\centering
\includegraphics[width=.5\textwidth]{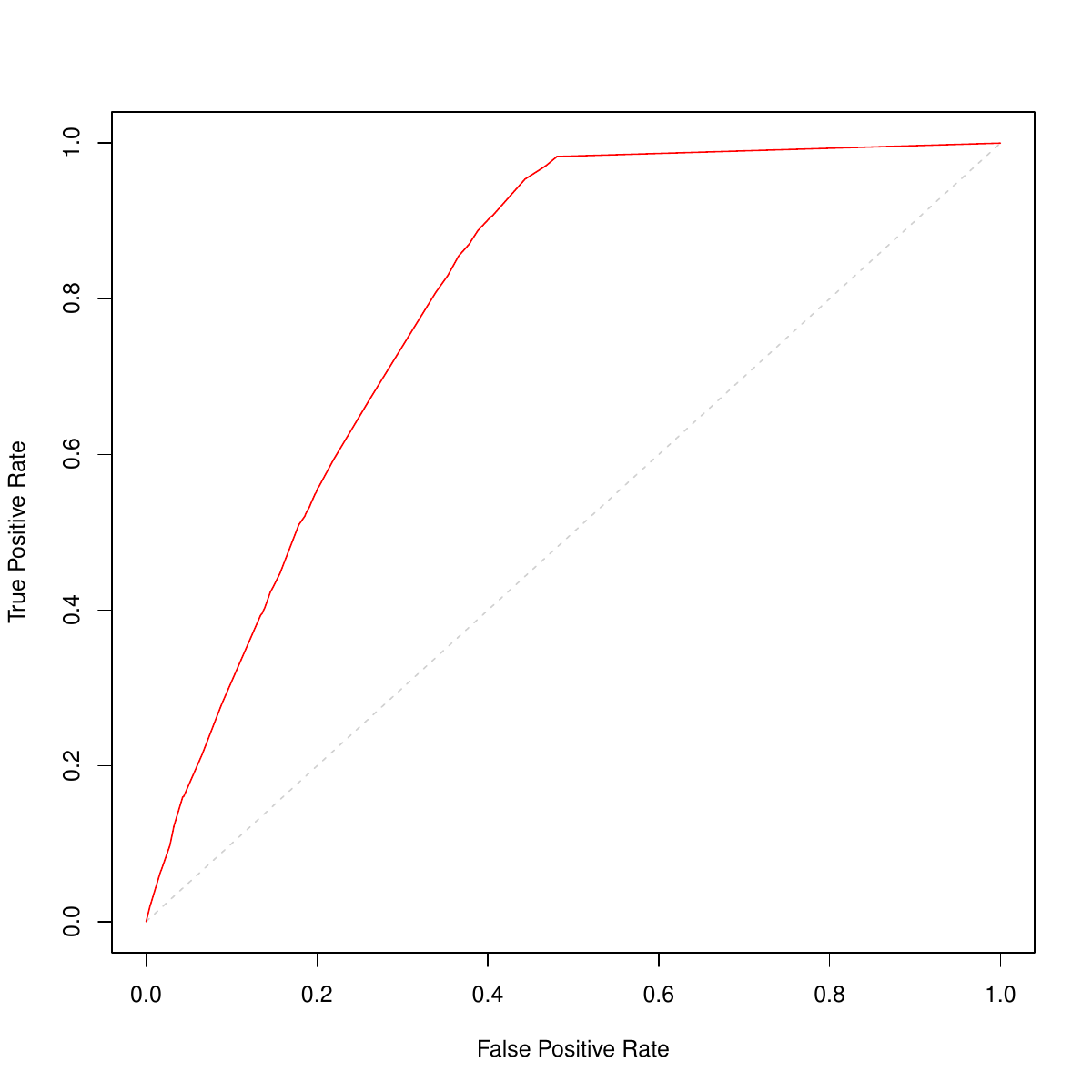} 
\caption{Receiver operating characteristic curve for a randomly selected validation subset with the IMDb dataset. }
\label{IMDbP}
\end{figure}

Finally, we tested the SGV algorithm using a real network constructed from a subset of the \emph{Internet Movie Database}. This network consists of $9,647$ vertices representing movies and $1,050,162$ edges indicating that two movies share at least one main cast member.  In this case the algorithm takes approximately eight hours and about 60 iterations (15 epochs) to reach convergence with $K=40$, $\omega=0.25$, $\kappa=0.6$, and $\tau=1$. As before, we select the best out of 32 runs of the algorithm.  The results suggest the existence of $K^\star=25$ different communities.  Interestingly, we observe an association between the resulting communities and the IMDb genre classification.  For example, a third of documentaries in the data are clustered together.  To evaluate the out-of-sample predictive performance of the model, Figure \ref{IMDbP} shows the receiver operating characteristic curve associated with a held out set that includes randomly selected 5\% of both positive and negative links, whose area under the curve is 0.8.

To further explore the performance of the SGV algorithm for a dataset with these dimensions, we created a simulated dataset with similar characteristics to the IMDb network.  Specifically, we take the recovered community structure and mean variational community parameters and set them as the ground truth.  Then, using these values, we randomly generate a new set of edges among the same $9,647$ vertices.  The resulting network consists of $1,051,101$ connections.  Again setting $K=40$, $\kappa=0.6$m and $\tau=1$, we run the algorithm 32 times for each of two different values of the proportion of subsampled vertices, $\omega=0.15$, and $\omega=0.25$.  The ARI corresponding to the highest achieving lower bound out of 32 runs are $0.5$ and $0.7$ respectively.  Hence, the algorithm recovers the underlying structure only partially and, not surprisingly, tends to perform better when a larger sub-batch is utilized.

\begin{figure}[!htb]
\centering
\includegraphics[width=.5\textwidth]{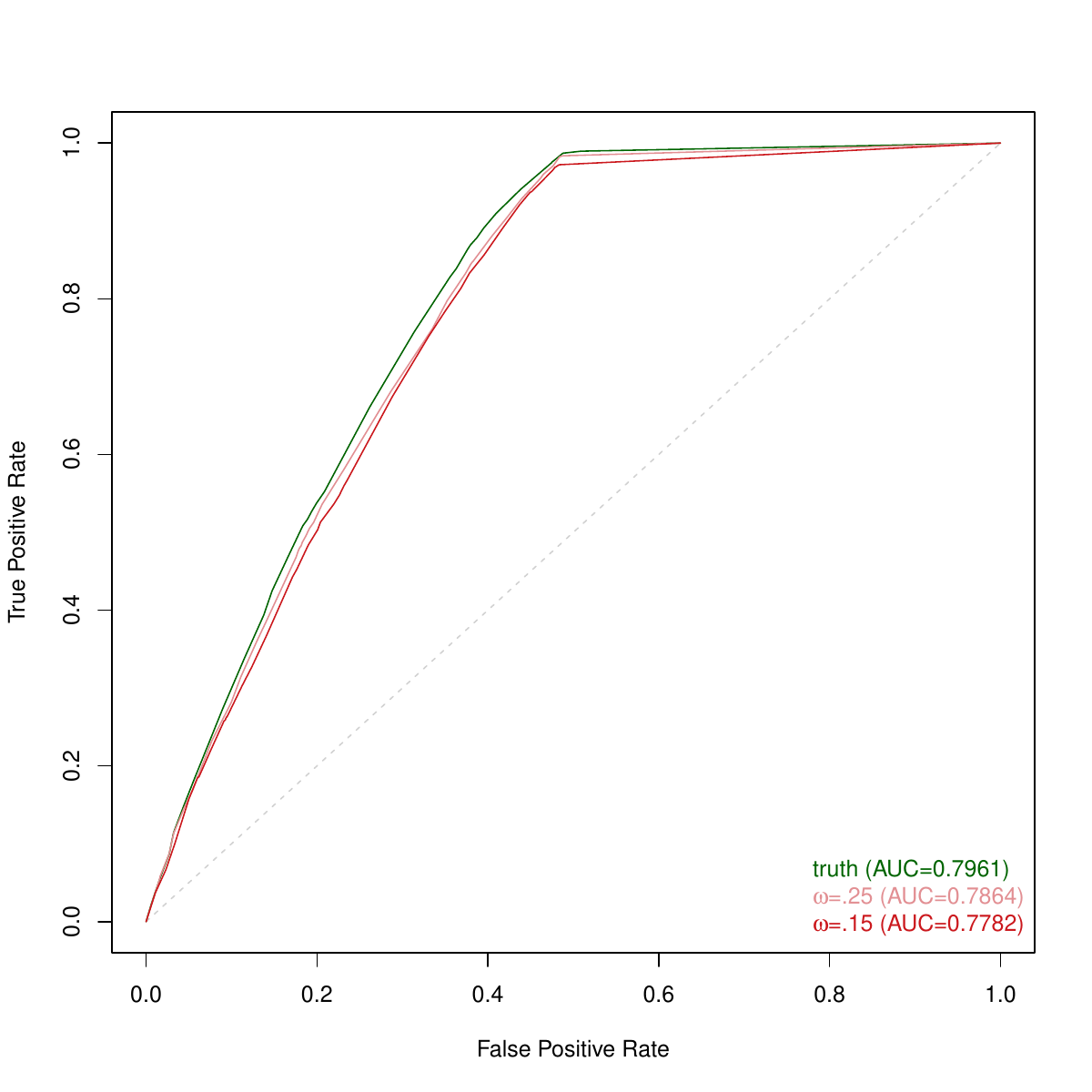} 
\caption{Receiver operating characteristic curves for a randomly selected validation subset with the simulated dataset.}
\label{SIMDbP}
\end{figure}

In addition to assessing the quality of the communities identified by the variational algorithm, we evaluate the predictive performance of the algorithm for both values of $\omega$.  Figure \ref{SIMDbP} presents ROC curves calculated by predicting a randomly selected subset of $25,000$ interactions in the network using the true model that generated the data, as well as both variational approximations.  Somewhat surprisingly, the predictive performance of the variational approximations is fairly close to prediction under the truth in both cases, with $\omega=0.25$ slightly outperforming $\omega=0.15$ in terms of the AUC.

\section{Discussion}\label{disc}

Our results show that the quality of the approximation provided by the SGV algorithm is highly dependent on the value of the tuning parameters $\tau,$ $\kappa$, and $\omega$.  Sensitivity to the value of $\omega$ was particularly striking.  Providing general guidance for the selection of these parameters is complicated as the optimal choice will typically depend on true number, size and interaction probabilities among structurally equivalent groups in the data.  However, our experiments suggest that subsamples of at least $25\%$ of the data might be required to obtain adequate results, therefore limiting the gains in computational efficiency that come from the use of SGV algorithms.

Our results also indicate that, while SGV approximations can perform well in terms of link prediction, their ability to accurately identify the underlying communities associated with a stochastic blockmodel is quite limited.  In hindsight, the nature of the algorithm should make this obvious:  The use of a mean-field approximation means that the true (posterior) matrix of pairwise co-clustering probabilities is being approximated by a rank-1 PARAFAC decomposition.  When the co-clustering probabilities are not close to the extremes (either 0s or 1s), this rank-1 representation will typically be too inflexible to provide a good approximation.  This suggests that the use of variational algorithms in general, and SGV algorithms in particular, in the context of community identification algorithms should be considered highly suspect.

\section*{Statements and Declarations}

The authors declare that they have no known competing financial interests or personal relationships that could have appeared to influence the work reported in this article.

\bibliography{references.bib}

\begin{thebibliography}{}

\bibitem[Agarwal et~al., 2017]{agarwal2017second}
Agarwal, N., Bullins, B., and Hazan, E. (2017).
\newblock Second-order stochastic optimization for machine learning in linear
  time.
\newblock {\em The Journal of Machine Learning Research}, 18(1):4148--4187.

\bibitem[Airoldi et~al., 2008]{Airoldietal08}
Airoldi, E.~M., Blei, D.~M., Fienberg, S.~E., and Xing, E.~P. (2008).
\newblock Mixed membership stochastic blockmodels.
\newblock {\em Journal of Machine Learning Research}, 9(Sep):1981--2014.

\bibitem[Antoniak, 1974]{Antoniak74}
Antoniak, C.~E. (1974).
\newblock Mixtures of dirichlet processes with applications to bayesian
  nonparametric problems.
\newblock {\em The annals of statistics}, pages 1152--1174.

\bibitem[Bender and Canfield, 1978]{Bender&Canfield78}
Bender, E.~A. and Canfield, E.~R. (1978).
\newblock The asymptotic number of labeled graphs with given degree sequences.
\newblock {\em Journal of Combinatorial Theory, Series A}, 24(3):296--307.

\bibitem[Bickel et~al., 2013]{Bickeletal13}
Bickel, P., Choi, D., Chang, X., and Zhang, H. (2013).
\newblock Asymptotic normality of maximum likelihood and its variational
  approximation for stochastic blockmodels.
\newblock {\em The Annals of Statistics}, pages 1922--1943.

\bibitem[Blei et~al., 2017]{blei2017variational}
Blei, D.~M., Kucukelbir, A., and McAuliffe, J.~D. (2017).
\newblock Variational inference: A review for statisticians.
\newblock {\em Journal of the American Statistical Association},
  112(518):859--877.

\bibitem[Boccaletti et~al., 2006]{Boccalettietal06}
Boccaletti, S., Latora, V., Moreno, Y., Chavez, M., and Hwang, D.-U. (2006).
\newblock Complex networks: Structure and dynamics.
\newblock {\em Physics reports}, 424(4):175--308.

\bibitem[Cao et~al., 2023]{cao2023variational}
Cao, J., Kang, M., Jimenez, F., Sang, H., Schaefer, F.~T., and Katzfuss, M.
  (2023).
\newblock Variational sparse inverse cholesky approximation for latent gaussian
  processes via double kullback-leibler minimization.
\newblock In {\em International Conference on Machine Learning}, pages
  3559--3576. PMLR.

\bibitem[Celisse et~al., 2012]{Celisseetal12}
Celisse, A., Daudin, J.-J., Pierre, L., et~al. (2012).
\newblock Consistency of maximum-likelihood and variational estimators in the
  stochastic block model.
\newblock {\em Electronic Journal of Statistics}, 6:1847--1899.

\bibitem[Chen et~al., 2014]{chen2014stochastic}
Chen, T., Fox, E., and Guestrin, C. (2014).
\newblock Stochastic gradient hamiltonian monte carlo.
\newblock In {\em International Conference on Machine Learning}, pages
  1683--1691.

\bibitem[Ewens, 1972]{Ewens72}
Ewens, W.~J. (1972).
\newblock The sampling theory of selectively neutral alleles.
\newblock {\em Theoretical population biology}, 3(1):87--112.

\bibitem[Fortunato, 2010]{Fortunato10}
Fortunato, S. (2010).
\newblock Community detection in graphs.
\newblock {\em Physics Reports}, 486(3):75--174.

\bibitem[Foulds et~al., 2013]{Fouldsetal13}
Foulds, J., Boyles, L., DuBois, C., Smyth, P., and Welling, M. (2013).
\newblock Stochastic collapsed variational bayesian inference for latent
  dirichlet allocation.
\newblock In {\em Proceedings of the 19th ACM SIGKDD international conference
  on Knowledge discovery and data mining}, pages 446--454. ACM.

\bibitem[Frank and Strauss, 1986]{Frank&Strauss86}
Frank, O. and Strauss, D. (1986).
\newblock Markov graphs.
\newblock {\em Journal of the american Statistical association},
  81(395):832--842.

\bibitem[Gamerman and Lopes, 2006]{gamerman2006markov}
Gamerman, D. and Lopes, H.~F. (2006).
\newblock {\em Markov chain Monte Carlo: stochastic simulation for Bayesian
  inference}.
\newblock Chapman and Hall/CRC.

\bibitem[Gelfand and Smith, 1990]{Gelfand&Smith90}
Gelfand, A.~E. and Smith, A.~F. (1990).
\newblock Sampling-based approaches to calculating marginal densities.
\newblock {\em Journal of the American statistical association},
  85(410):398--409.

\bibitem[Geman and Geman, 1984]{Geman&Geman84}
Geman, S. and Geman, D. (1984).
\newblock Stochastic relaxation, gibbs distributions, and the bayesian
  restoration of images.
\newblock {\em Pattern Analysis and Machine Intelligence, IEEE Transactions
  on}, 6(6):721--741.

\bibitem[Goldenberg et~al., 2010]{Goldenbergetal10}
Goldenberg, A., Zheng, A.~X., Fienberg, S.~E., and Airoldi, E.~M. (2010).
\newblock A survey of statistical network models.
\newblock {\em Foundations and Trends in Machine Learning}, 2(2):129--233.

\bibitem[Gopalan et~al., 2012]{Gopalanetal12}
Gopalan, P.~K., Gerrish, S., Freedman, M., Blei, D.~M., and Mimno, D.~M.
  (2012).
\newblock Scalable inference of overlapping communities.
\newblock In {\em Advances in Neural Information Processing Systems}, pages
  2249--2257.

\bibitem[Graham, 1994]{graham1994concrete}
Graham, R.~L. (1994).
\newblock {\em Concrete mathematics: a foundation for computer science}.
\newblock Pearson Education India.

\bibitem[Hastings, 1970]{Hastings70}
Hastings, W.~K. (1970).
\newblock Monte carlo sampling methods using markov chains and their
  applications.
\newblock {\em Biometrika}, 57(1):97--109.

\bibitem[Hoff, 2005]{hoff2005bilinear}
Hoff, P.~D. (2005).
\newblock Bilinear mixed-effects models for dyadic data.
\newblock {\em Journal of the {A}merican {S}tatistical {A}ssociation},
  100(469):286--295.

\bibitem[Hoff, 2008]{hoff-2008}
Hoff, P.~D. (2008).
\newblock Modeling homophily and stochastic equivalence in symmetric relational
  data.
\newblock In {\em Advances in Neural Information Processing Systems}, pages
  657--664.

\bibitem[Hoff, 2009]{hoff-2009-multiplicative}
Hoff, P.~D. (2009).
\newblock Multiplicative latent factor models for description and prediction of
  social networks.
\newblock {\em Computational and Mathematical Organization Theory},
  15(4):261--272.

\bibitem[Hoff et~al., 2002]{Hoffetal02}
Hoff, P.~D., Raftery, A.~E., and Handcock, M.~S. (2002).
\newblock Latent space approaches to social network analysis.
\newblock {\em Journal of the american Statistical association},
  97(460):1090--1098.

\bibitem[Hoffman and Blei, 2015]{hoffman2015structured}
Hoffman, M.~D. and Blei, D.~M. (2015).
\newblock Structured stochastic variational inference.
\newblock In {\em Artificial Intelligence and Statistics}, pages 361--369.

\bibitem[Hoffman et~al., 2013]{Hoffmanetal13}
Hoffman, M.~D., Blei, D.~M., Wang, C., and Paisley, J. (2013).
\newblock Stochastic variational inference.
\newblock {\em The Journal of Machine Learning Research}, 14(1):1303--1347.

\bibitem[Holland et~al., 1983]{Hollandetal83}
Holland, P.~W., Laskey, K.~B., and Leinhardt, S. (1983).
\newblock Stochastic blockmodels: First steps.
\newblock {\em Social networks}, 5(2):109--137.

\bibitem[Hubert and Arabie, 1985]{Hubert&Arabie85}
Hubert, L. and Arabie, P. (1985).
\newblock Comparing partitions.
\newblock {\em Journal of classification}, 2(1):193--218.

\bibitem[Ishwaran and Zarepour, 2000]{Ishwaran&Zarepour00}
Ishwaran, H. and Zarepour, M. (2000).
\newblock Markov chain monte carlo in approximate dirichlet and beta
  two-parameter process hierarchical models.
\newblock {\em Biometrika}, 87(2):371--390.

\bibitem[Jaakkola and Jordan, 2000]{Jaakkola&Jordan00}
Jaakkola, T.~S. and Jordan, M.~I. (2000).
\newblock Bayesian parameter estimation via variational methods.
\newblock {\em Statistics and Computing}, 10(1):25--37.

\bibitem[Jordan et~al., 1999]{Jordanetal99}
Jordan, M.~I., Ghahramani, Z., Jaakkola, T.~S., and Saul, L.~K. (1999).
\newblock An introduction to variational methods for graphical models.
\newblock {\em Machine learning}, 37(2):183--233.

\bibitem[Kemp et~al., 2006]{Kempetal06}
Kemp, C., Tenenbaum, J.~B., Griffiths, T.~L., Yamada, T., and Ueda, N. (2006).
\newblock Learning systems of concepts with an infinite relational model.
\newblock In {\em AAAI}, volume~3, page~5.

\bibitem[Kiefer and Wolfowitz, 1952]{kiefer1952stochastic}
Kiefer, J. and Wolfowitz, J. (1952).
\newblock Stochastic estimation of the maximum of a regression function.
\newblock {\em The Annals of Mathematical Statistics}, pages 462--466.

\bibitem[Kim et~al., 2013]{Kimetal13}
Kim, D.~I., Gopalan, P.~K., Blei, D., and Sudderth, E. (2013).
\newblock Efficient online inference for bayesian nonparametric relational
  models.
\newblock In {\em Advances in Neural Information Processing Systems}, pages
  962--970.

\bibitem[Kim et~al., 2022]{kim2022markov}
Kim, K., Oh, J., Gardner, J., Dieng, A.~B., and Kim, H. (2022).
\newblock Markov chain score ascent: A unifying framework of variational
  inference with markovian gradients.
\newblock {\em Advances in Neural Information Processing Systems},
  35:34802--34816.

\bibitem[Kingma and Welling, 2014]{kingma2014stochastic}
Kingma, D.~P. and Welling, M. (2014).
\newblock Stochastic gradient vb and the variational auto-encoder.
\newblock In {\em Second International Conference on Learning Representations,
  ICLR}.

\bibitem[Knowles, 2015]{knowles2015stochastic}
Knowles, D.~A. (2015).
\newblock Stochastic gradient variational bayes for gamma approximating
  distributions.
\newblock {\em arXiv preprint arXiv:1509.01631}.

\bibitem[Kurihara et~al., 2007]{Kuriharaetal07}
Kurihara, K., Welling, M., and Teh, Y.~W. (2007).
\newblock Collapsed variational dirichlet process mixture models.
\newblock In {\em IJCAI}, volume~7, pages 2796--2801.

\bibitem[Latouche et~al., 2012]{latouche2012variational}
Latouche, P., Birmele, E., and Ambroise, C. (2012).
\newblock Variational bayesian inference and complexity control for stochastic
  block models.
\newblock {\em Statistical Modelling}, 12(1):93--115.

\bibitem[Lau and Green, 2007]{Lau&Green07}
Lau, J.~W. and Green, P.~J. (2007).
\newblock Bayesian model-based clustering procedures.
\newblock {\em Journal of Computational and Graphical Statistics},
  16(3):526--558.

\bibitem[Lesne, 2014]{lesne2014shannon}
Lesne, A. (2014).
\newblock Shannon entropy: a rigorous notion at the crossroads between
  probability, information theory, dynamical systems and statistical physics.
\newblock {\em Mathematical Structures in Computer Science}, 24(3):e240311.

\bibitem[Li et~al., 2024]{li2024comprehensive}
Li, J., Lai, S., Shuai, Z., Tan, Y., Jia, Y., Yu, M., Song, Z., Peng, X., Xu,
  Z., Ni, Y., Qiu, H., Yang, J., Liu, Y., and Lu, Y. (2024).
\newblock A comprehensive review of community detection in graphs.
\newblock {\em Neurocomputing}, 600:128169.

\bibitem[Ma et~al., 2015]{ma2015complete}
Ma, Y.-A., Chen, T., and Fox, E. (2015).
\newblock A complete recipe for stochastic gradient mcmc.
\newblock In {\em Advances in Neural Information Processing Systems}, pages
  2917--2925.

\bibitem[Mandt et~al., 2016]{mandt2016variational}
Mandt, S., Hoffman, M., and Blei, D. (2016).
\newblock A variational analysis of stochastic gradient algorithms.
\newblock In {\em International conference on machine learning}, pages
  354--363. PMLR.

\bibitem[Mariadassou et~al., 2010]{mariadassou2010uncovering}
Mariadassou, M., Robin, S., and Vacher, C. (2010).
\newblock Uncovering latent structure in valued graphs: a variational approach.
\newblock {\em The Annals of Applied Statistics}, pages 715--742.

\bibitem[McNamara et~al., 2024]{mcnamara2024sequential}
McNamara, D., Loper, J., and Regier, J. (2024).
\newblock Sequential monte carlo for inclusive kl minimization in amortized
  variational inference.
\newblock In {\em International Conference on Artificial Intelligence and
  Statistics}, pages 4312--4320. PMLR.

\bibitem[Metropolis et~al., 1953]{Metropolisetal53}
Metropolis, N., Rosenbluth, A.~W., Rosenbluth, M.~N., Teller, A.~H., and
  Teller, E. (1953).
\newblock Equation of state calculations by fast computing machines.
\newblock {\em The journal of chemical physics}, 21(6):1087--1092.

\bibitem[Neal, 2000]{Neal00}
Neal, R.~M. (2000).
\newblock Markov chain sampling methods for dirichlet process mixture models.
\newblock {\em Journal of computational and graphical statistics},
  9(2):249--265.

\bibitem[Newman, 2003]{Newman03}
Newman, M.~E. (2003).
\newblock The structure and function of complex networks.
\newblock {\em SIAM review}, 45(2):167--256.

\bibitem[Newman, 2004]{Newman04}
Newman, M.~E. (2004).
\newblock Detecting community structure in networks.
\newblock {\em The European Physical Journal B-Condensed Matter and Complex
  Systems}, 38(2):321--330.

\bibitem[Newman, 2006]{Newman06}
Newman, M.~E. (2006).
\newblock Finding community structure in networks using the eigenvectors of
  matrices.
\newblock {\em Physical review E}, 74(3):036104.

\bibitem[Polatkan et~al., 2015]{Polatkanetal15}
Polatkan, G., Zhou, M., Carin, L., Blei, D., and Daubechies, I. (2015).
\newblock A bayesian nonparametric approach to image super-resolution.
\newblock {\em IEEE transactions on pattern analysis and machine intelligence},
  37(2):346--358.

\bibitem[Porter et~al., 2009]{Porteretal09}
Porter, M.~A., Onnela, J.-P., and Mucha, P.~J. (2009).
\newblock Communities in networks.
\newblock {\em Notices of the AMS}, 56(9):1082--1097.

\bibitem[Robbins and Monro, 1951]{Robbins&Monro51}
Robbins, H. and Monro, S. (1951).
\newblock A stochastic approximation method.
\newblock {\em The annals of mathematical statistics}, pages 400--407.

\bibitem[Robert and Casella, 2005]{robertcasella2004monte}
Robert, C.~P. and Casella, G. (2005).
\newblock {\em Monte Carlo Statistical Methods}.
\newblock Springer, second edition.

\bibitem[Ruder, 2016]{DBLP:journals/corr/Ruder16}
Ruder, S. (2016).
\newblock An overview of gradient descent optimization algorithms.
\newblock {\em CoRR}, abs/1609.04747.

\bibitem[Saul et~al., 1996]{Sauletal96}
Saul, L.~K., Jaakkola, T., and Jordan, M.~I. (1996).
\newblock Mean field theory for sigmoid belief networks.
\newblock {\em Journal of Artificial Intelligence Research}, 4(61):76.

\bibitem[Schaeffer, 2007]{Schaeffer07}
Schaeffer, S.~E. (2007).
\newblock Graph clustering.
\newblock {\em Computer Science Review}, 1(1):27--64.

\bibitem[Snijders and Nowicki, 1997]{snijders1997estimation}
Snijders, T.~A. and Nowicki, K. (1997).
\newblock Estimation and prediction for stochastic blockmodels for graphs with
  latent block structure.
\newblock {\em Journal of classification}, 14(1):75--100.

\bibitem[Tabouy et~al., 2020]{tabouy2020variational}
Tabouy, T., Barbillon, P., and Chiquet, J. (2020).
\newblock Variational inference for stochastic block models from sampled data.
\newblock {\em Journal of the American Statistical Association},
  115(529):455--466.

\bibitem[Tran et~al., 2021]{tran2021variational}
Tran, M.-N., Nguyen, D.~H., and Nguyen, D. (2021).
\newblock Variational bayes on manifolds.
\newblock {\em Statistics and Computing}, 31:1--17.

\bibitem[Wang and Titterington, 2004a]{wang2004convergence}
Wang, B. and Titterington, D.~M. (2004a).
\newblock Convergence and asymptotic normality of variational bayesian
  approximations for exponential family models with missing values.
\newblock In {\em Proceedings of the 20th conference on Uncertainty in
  artificial intelligence}, pages 577--584. AUAI Press.

\bibitem[Wang and Titterington, 2004b]{wang2004lack}
Wang, B. and Titterington, D.~M. (2004b).
\newblock Lack of consistency of mean field and variational bayes
  approximations for state space models.
\newblock {\em Neural Processing Letters}, 20(3):151--170.

\bibitem[Wang and Wong, 1987]{wang1987stochastic}
Wang, Y.~J. and Wong, G.~Y. (1987).
\newblock Stochastic blockmodels for directed graphs.
\newblock {\em Journal of the American Statistical Association}, 82(397):8--19.

\bibitem[Welling and Teh, 2011]{welling2011bayesian}
Welling, M. and Teh, Y.~W. (2011).
\newblock Bayesian learning via stochastic gradient langevin dynamics.
\newblock In {\em Proceedings of the 28th International Conference on Machine
  Learning (ICML-11)}, pages 681--688.

\end{thebibliography}
\bibliographystyle{apalike}



\end{document}